\documentclass[12pt,preprint]{aastex}

\usepackage{natbib}

\begin{document}


\title{Production of Light Element Primary Process nuclei in
  neutrino-driven winds}


\author{A.~Arcones} \affil{Institut f\"ur Kernphysik, TU~Darmstadt,
  Schlossgartenstr.~9, 64289 Darmstadt, Germany} \affil{GSI
  Helmholtzzentrum f\"ur Schwerionenforschung, Planckstr. 1, 64291
  Darmstadt, Germany} \affil{Department of Physics, University of
  Basel, Klingelbergstra{\ss}e 82, 4056, Basel, Switzerland}

\author{F.~Montes} 
 \affil{Joint Institute for Nuclear Astrophysics,  http://www.jinaweb.org}
 \affil{National Superconducting Cyclotron  Laboratory, Michigan State University, East Lansing, MI 48824, USA}

\begin{abstract}
  We present first comparisons between Light Element Primary Process
  (LEPP) abundances observed in some ultra metal poor (UMP) stars and
  nucleosynthesis calculations based on long-time hydrodynamical
  simulations of core-collapse supernovae and their neutrino-driven
  wind. UMP star observations indicate Z$\ge$38 elements include the
  contributions of at least two nucleosynthesis components: r-process
  nuclei that are synthesized by rapid neutron capture in a yet
  unknown site and LEPP elements (mainly Sr, Y, Zr). We show that
  neutrino-driven wind simulations can explain the observed LEPP
  pattern. We explore in detail the sensitivity of the calculated
  abundances to the electron fraction, which is a key nucleosynthesis
  parameter but poorly known due to uncertainties in neutrino
  interactions and transport. Our results show that the observed LEPP
  pattern can be reproduced in proton- and neutron-rich winds.
\end{abstract}


\section{Introduction}
\label{sec:introduction}

Elements heavier than iron are mainly produced by the slow (s-) and
rapid (r-) neutron capture processes. In contrast to the s-process,
for which the astrophysical environments are identified
~\cite[]{Busso.etal:1999, Straniero.etal:2006}, the r-process site(s)
remains unknown~\cite[]{arnould.goriely.takahashi:2007}. There are
some indications that more than one component or site contributes to
the abundances of the so called r-process
elements~\cite[]{Wasserburg.etal:1996,Qian.Wasserburg:2001,
  Truran.Cowan.ea:2002, Travaglio.Gallino.ea:2004, Aoki.etal:2005,
  Otsuki.etal:2006,Montes.etal:2007}. Most of the recent progress in
understanding the origin of elements commonly associated with the
r-process is due to observations of ultra metal-poor (UMP) stars
~\cite[see][for a recent review]{Sneden.etal:2008}.  The elemental
abundances observed in the atmosphere of these very old stars come
from a few events.  These stars generally present a robust pattern for
``heavy'' elements $56<Z<83$, in agreement with the expected
contribution of the r-process to the solar system, but show some
scatter for ``light'' elements $Z<47$
~\cite[]{Johnson.Bolte:2002,Sneden.etal:2008}. This suggests that at least two
types of primary nucleosynthesis events contribute to the abundances
of those elements.

The first indication for these two nucleosynthesis components came
from meteoritic data on $^{129}$I and $^{182}$Hf
~\cite[]{Wasserburg.etal:1996}. The idea of the different components
was later extended to account for all abundances in metal-poor stars
in the phenomenological model of ~\cite{Qian.Wasserburg:2001,
  Qian.Wasserburg:2007}.  They argued that supernovae from low-mass
progenitors with $8 M_\odot< M < 12 M_\odot$ (H-source in their
terminology) lead to all ``heavy'' and some ``light'' elements, and
that explosions of more massive progenitors, $12 M_\odot< M < 25
M_\odot$ (L-source), contribute to the remaining light $A<130$
elements.  A similar conclusion was reached independently by
~\cite{Montes.etal:2007} using a slightly different pattern for the
L-source (LEPP source in their terminology).

The abundances of star HD~122563~\cite[]{Westin.etal:2000} were used
to build an L-source pattern. This star shows mainly elements produced
in the L-source with small contamination from the H-source,
characterized by the abundances observed in
CS~22892-052~\cite[]{Sneden.etal:2003}.  While
~\cite{Qian.Wasserburg:2007} assumed that all material above $Z=56$ in
HD~122563 was produced by the H-source, ~\cite{Montes.etal:2007} used
a smaller H-source contribution responsible for all the abundance of
elements heavier than $Z=62$.  ~\cite{Montes.etal:2007} discussed the
assumption that HD~122563 is representative of the L-source by showing
a hint of robustness of the observed pattern of UMP stars. However,
more observations of L-source enriched (or r-process poor) UMP stars
are needed before a definitive conclusion is reached. In this paper,
the LEPP (or L-source) pattern was taken from
\cite{Qian.Wasserburg:2008} based on the abundances of
\cite{Honda.etal:2006}. It covers elements in the range $38<Z<47$ and
it is assumed to be robust.

More recently,~\cite{Qian.Wasserburg:2008} updated their
two-component model to include a third component producing Fe but not
$Z\ge38$ elements.  The need for an additional component resulted from
using newer higher-resolution data that included many stars with
$\textrm{[Fe/H]}\le -3$ where the amount of Fe (produced originally
only in the L-source) relative to Sr (produced in both L- and H-
sources) was too large compared to the yields based on the
two-component model.

The process leading to elements with $A<130$ (L-source) has been
called in the literature the weak r-process
~\cite[]{Truran.Cowan:2000}, charged-particle reaction (CPR) process
~\cite[]{Woosley.Hoffman:1992, Freiburghaus.Rembges.ea:1999,
  Qian.Wasserburg:2007}, and Lighter Element Primary Process (LEPP)
~\cite[]{Travaglio.Gallino.ea:2004, Montes.etal:2007}. In this paper,
we refer to this process as LEPP because it does not make any
reference to the specific nuclear reactions or astrophysical
environment. The term LEPP was first introduced in
~\cite{Travaglio.Gallino.ea:2004} which used a galactic chemical
evolution model to search for the astrophysical environments producing
the elements Sr, Y, and Zr. Using their s-process model and standard
r-process contributions, they found that non-negligible abundances of
several isotopes ($^{86}$Sr, $^{93}$Nb, $^{96}$Mo, $^{100}$Ru,
$^{104}$Pd, $^{110}$Cd) were still
unexplained. \cite{Montes.etal:2007} showed that the LEPP elemental
abundances that result from using the
~\cite{Travaglio.Gallino.ea:2004} model are in agreement, within the
observational error bars, with the modified abundances of HD~122563
(Fig.~5 in ~\cite{Montes.etal:2007}). The contribution of the LEPP to
the solar abundances depends on the s-process model. If further
refinements to s-process models eliminate or change the need for a
LEPP contribution to the solar system abundances, the apparent
agreement between the ``solar'' LEPP and the UMP observations would
have been coincidental. In that case the LEPP may only contribute to
the abundances of a few metal-poor stars.

After the initial success of ~\cite{Woosley.Wilson.ea:1994} in
reproducing observed r-process abundances\footnote{Recent work of
  ~\cite{Roberts.etal:2010} suggests that the high wind entropies
  obtained at that time could have been due to problems with the
  equation of state.}, core-collapse supernovae and the subsequent
neutrino-driven winds became one of the most promising candidates for
the production of r-process elements because their extreme explosive
conditions are very close to the ones needed for the r-process
~\cite[see e.g.,][]{Hoffman97, Thompson.Burrows.Meyer:2001,
  Otsuki.Tagoshi.ea:2000}. Moreover, galactic chemical evolution
models favor core-collapse supernovae, since they occur early and
frequently enough to account for the abundances observed in old halo
stars and in the solar system ~\cite[]{Ishimaru.Wanajo:1999,
  Ishimaru.etal:2004}.  Although the necessary conditions to produce
heavy elements ($A>130$) are identified ~\cite[]{Meyer92} (high
entropies, low electron fractions, and short expansion timescales),
these are not found in the most recent long-time supernova simulations
~\cite[]{Pruet.Hoffman.ea:2006, arcones.janka.scheck:2007,
  Wanajo.Nomoto.ea:2009, Fischer.etal:2010,Huedepohl.ea:2010}.

In this paper we investigate the possibility of producing the LEPP
elements in neutrino-driven winds, as suggested by
~\cite{Qian.Wasserburg:2007, Qian.Wasserburg:2008}. The
nucleosynthesis calculations (Sect.~\ref{sec:netw}) are based on the
spherically symmetric simulations of core-collapse supernovae of
~\cite{arcones.janka.scheck:2007}, where the evolution of the
neutrino-driven wind was followed during several seconds
(Sect.~\ref{sec:sn_wind}). Since the neutrino transport in these
simulations is only approximate
~\cite[]{Scheck.Kifonidis.Janka.Mueller:2006}, the electron fraction
is expected to have some uncertainty as explained in
Sect.~\ref{sec:windparameters}. A systematic study of the impact of
the wind electron fraction on the nucleosynthesis is included in
Sect.~\ref{sec:yevar}. We show that the nucleosynthesis based on these
simulations do not yield heavy r-process nuclei ($A>130$) but that
LEPP elements can be produced in the current neutrino-driven wind
models. The different conditions to produce the observed LEPP pattern
are investigated in Sect.~\ref{sec:lepp}. Finally, we conclude and
summarize Sect.~\ref{sec:conclusions}.

\section{Nucleosynthesis in neutrino-driven winds}
\label{sec:nuc}
When a supernova explodes, matter surrounding the proto-neutron star
is heated by neutrinos and expands very fast reaching sometimes even
supersonic velocity ~\cite[]{duncan.shapiro.wasserman:1986,
  Thompson.Burrows.Meyer:2001}. This is known as the neutrino-driven
wind and can become neutron or proton rich. The electron fraction
$Y_e$ is determined by the uncertain neutrino properties (energy and
luminosity) in the region where neutrinos decouple from matter.  This
matter near the neutron star consist mainly of neutrons and protons
due to the high temperatures in this region.  When a mass element
expands, its temperature decreases and neutrons and protons recombine
to form $\alpha$-particles. The density decreases but as the
triple-alpha reaction combined with different $\alpha$ capture
reactions are still occurring, heavy seed nuclei may
form~\cite[]{Woosley.Hoffman:1992, Witti.Janka.Takahashi:1994}. The
evolution once the $\alpha$-particles start forming heavier nuclei
depends on the electron fraction and it will be discussed in detail in
Sect.~\ref{sec:yevar}.

\subsection{Wind electron fraction}
\label{sec:windparameters}
The evolution of the nascent neutron star and the properties of
neutrinos emitted from its surface have a direct impact on the wind
parameters that are relevant for nucleosynthesis: entropy ($S$),
expansion timescale ($\tau=r/v|_{T\sim0.5\mathrm{MeV}}$), and electron
fraction.  $S$ and $\tau$ depend on neutrino energy, neutrino
luminosity, and on the evolution of the proto-neutron star as
~\cite[]{Qian.Woosley:1996}:
\begin{eqnarray}
  S   & \propto & L^{-1/6} \epsilon^{-1/3} R_{\mathrm{ns}}^{-2/3} M_{\mathrm{ns}}\, ,\label{eq:QW_S} \\
  \tau & \propto & L^{-1} \epsilon^{-2} R_{\mathrm{ns}} M_{\mathrm{ns}} \, ,         \label{eq:QW_tau}
\end{eqnarray}
where $L$, $\epsilon$ are the neutrino luminosity and mean energy,
respectively, and the dependence on the neutron star properties enters
through its radius ($R_{\mathrm{ns}}$) and mass
($M_{\mathrm{ns}}$). The evolution of these quantities is mainly given
by the equation of state (EoS) at nuclear densities, which is
specially uncertain at high densities, and by the amount of matter
accreted onto the proto-neutron star, which is higher for more massive
stellar progenitors. Eqs.~(\ref{eq:QW_S})-(\ref{eq:QW_tau}) indicate
that different progenitors with the same neutron star evolution lead
to similar wind properties.  When neutrino luminosities and energies
decrease, the entropy slightly increases favoring the production of
heavier elements, but at the same time the expansion timescale
increases which counteracts and disfavors their production. Therefore,
during the wind phase, relatively small changes in the nucleosynthesis
can generally be expected when the absolute values of neutrino
properties (i.e., $L_{\nu_e}+L_{\bar{\nu}_e}$ and
$\epsilon_{\nu_e}+\epsilon_{\bar{\nu}_e}$) are changed.  After the
initial fast contraction of the proto-neutron star, its radius and
mass changes only slightly for $\gtrsim$1~s after bounce. The main
change to $S$ and $\tau$ is then the result of the neutrino luminosity
variation as a function of time.

\begin{figure}[!htb]
  \includegraphics[width=8cm]{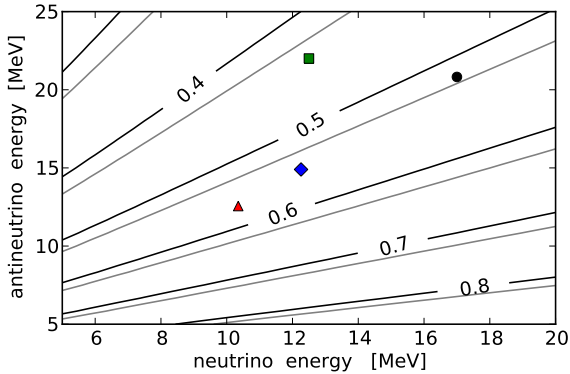}
  \caption{Electron fraction based on Eq.~(\ref{eq:ye}) for different
    neutrino (x-axis) and antineutrino (y-axis) energies. The black
    contours correspond to $L_{\bar{\nu}_e}/L_{\nu_e}=1$ and the grey
    contours to $L_{\bar{\nu}_e}/L_{\nu_e}=1.1$. The symbols show the
    electron neutrino and antineutrino energies ($\epsilon_{\nu_e}
    \approx 4.1 kT_{\nu_e}$) for different supernova models reported
    in the literature: square for~\cite{Woosley.Wilson.ea:1994},
    circle for model M15-l1-r6 of ~\cite{arcones.janka.scheck:2007}
    (note that these models do not include red shift corrections which
    would reduce the neutrino energies close to the neutron star),
    triangle for a 10~$M_{\odot}$ progenitor
    of~~\cite{Fischer.etal:2010}, and diamond
    for~\cite{Huedepohl.ea:2010}, all at 10~s after bounce.  Adapted
    from ~\cite{Qian.Woosley:1996}.}
  \label{fig:neut2col}
\end{figure}

The electron fraction is, contrary to $S$ and $\tau$, determined by
the relative variations of electron neutrinos and antineutrinos
properties. The exact value of neutrino energies and luminosities
depend on the accuracy of the supernova neutrino transport
calculations and on details of neutrino interactions at high densities
~\cite[]{Rampp.Janka:2000, Mezzacappa.Liebendoerfer.ea:2001}. At the
high temperatures ($T \gtrsim 10$~GK) where the nucleosynthesis
calculations are started, nuclei are photo-dissociated into neutrons
and protons and their abundances are given by $Y_p=Y_e$ and
$Y_n=1-Y_e$, respectively. The electron fraction is determined by
charged-current reactions between neutrinos and nucleons and their
inverse reactions:
\begin{eqnarray}
  \nu_e + n &\longleftrightarrow & p + e        \, ,   \label{eq:nuen_n}\\
  \bar{\nu}_e + p &\longleftrightarrow & n + e^+\, .    \label{eq:nuea_p}
\end{eqnarray}
The evolution of the electron fraction is consequently given by:
\begin{equation}
  \frac{\mathrm{d}Y_e}{\mathrm{d}t} =
  \lambda_{\nu_e n}Y_n -\lambda_{p e}Y_p -\lambda_{\bar{\nu}_e p}Y_p +\lambda_{n e^+}Y_n\, ,  
  \label{eq:dye_dt}
\end{equation}
where $\lambda_i$ are the reaction rates for the forward and backward
reactions in Eqs.~(\ref{eq:nuen_n})-(\ref{eq:nuea_p}). Neutrino
emission reactions are negligible in the wind because their rates
rapidly drop with temperature ($\lambda_{p e/ n e^+} \propto
T^5$). Following ~\cite{Qian.Woosley:1996}, the equilibrium initial
$Y_e$ in the wind is given by
\begin{equation}
  Y_e=\frac{\lambda_{\nu_e,n}}{\lambda_{\nu_e,n}+\lambda_{\bar{\nu}_e,p}}=
  \left[ 
    1+\frac{L_{\bar{\nu}_e}}{L_{\nu_e}}
    \frac{\epsilon_{\bar{\nu}_e} -2\Delta+1.2\Delta^2/\epsilon_{\bar{\nu}_e}}
    {\epsilon_{\nu_e} +2\Delta+1.2\Delta^2/\epsilon_{\nu_e}}
  \right]^{-1} \, .
  \label{eq:ye}
\end{equation}
Here the neutrino absorption rates are used in a simple form
~\cite[]{Qian.Woosley:1996} without considering weak magnetism and
recoil corrections ~\cite[]{Horowitz.Li:1999}. $L_{\nu_e}$ is the
neutrino luminosity, $\epsilon_{\nu_e} = \langle \varepsilon^2\rangle
/ \langle \varepsilon\rangle \approx 4.1 kT_{\nu_e}$ is the ratio of
mean squared neutrino energy ($\langle \varepsilon^2\rangle$) and mean
neutrino energy ($\langle \varepsilon\rangle$), and $kT_{\nu_e}$ is
the neutrino temperature in MeV (and similarly for the antineutrinos),
$\Delta$ denotes the neutron-proton mass difference. This simple
expression gives only an estimate of the $Y_e$ value found by
supernova neutrino transport calculations~\cite[]{Fischer.etal:2010}.
The neutrino two-color plot shown in Fig.~\ref{fig:neut2col}
illustrates the dependence of the $Y_e$ on the neutrino energies and
luminosities using Eq.~(\ref{eq:ye}). It also shows the (anti)neutrino
energies obtained 10~s after bounce in wind simulations reported in
the literature. The most recent long-time supernova
simulations~\cite[]{Huedepohl.ea:2010,Fischer.etal:2010} obtain lower
antineutrino energies compared to the earlier models, leading to
proton-rich conditions in the wind.

Neutrino energies depend on the temperature of the medium in the
region where neutrinos decouple from matter. This region is known as
the neutrinosphere and its location is different for each neutrino
flavor and energy.  Inside of the neutrinosphere, neutrinos are in the
thermal equilibrium by charged-current reactions
(Eqs.~(\ref{eq:nuen_n})-(\ref{eq:nuea_p})).  Outside of it, neutrinos
escape while their temperature stays almost constant and approximately
equal to the temperature at their neutrinosphere.  Because neutrons
are more abundant than protons in the neutron star, electron neutrinos
continue interacting to larger radii and thus to lower temperatures
than antineutrinos ($\varepsilon_{\bar{\nu}_e} \gtrsim
\varepsilon_{\nu_e}$). The $\mu$ and $\tau$ (anti)neutrinos interact
only via neutral-current reactions and decouple at smaller radii,
therefore their energies are larger.  During the first seconds after
the explosion, the proto-neutron star deleptonizes and the amount of
protons in the outer layers decreases.  The electron antineutrino
energies are thus expected to get higher than the electron neutrino
energies leading to an increase of the ratio
$\varepsilon_{\bar{\nu}_e} / \varepsilon_{\nu_e}$. New hydrodynamical
simulations including detailed neutrino transport
~\cite[e.g.][]{Huedepohl.ea:2010} show that this simple picture is not
valid as the spectra of electron neutrinos and antineutrinos are
rather similar, probably due to neutral-current reactions that act in
a similar way on all neutrino flavors and become more important as the
neutron star cools ~\cite[]{Huedepohl.ea:2010,
  arcones.matinezpinedo.etal:2008}. If electron neutrino and
antineutrino energies are similar, the neutron-to-proton mass
difference favors a value of $Y_e > 0.5$ (as shown in
Fig.~\ref{fig:neut2col}) by reducing the antineutrino absorption rate
relative to the neutrino rate.

It should be noted that some uncertainty remain in the calculation of
the neutrino spectra. At the high temperatures and densities of the
outer layers of the proto-neutron star, there may also be a
non-negligible abundance of light nuclei: $^2$H, $^3$H, and $^3$He
~\cite[]{Arcones.etal:2010}. These details in the composition are not
included in state-of-the-art supernova simulations, although neutrino
interactions with these light nuclei have an impact on the neutrino
spectra and thus on the
$Y_e$~\cite[]{arcones.matinezpinedo.etal:2008}.

\subsection{Wind simulations}
\label{sec:sn_wind}
In order to accurately calculate the integrated nucleosynthesis in
neutrino-driven winds, the evolution of the supernova ejecta has to be
followed for several seconds using hydrodynamical simulations. Such
long-term modeling is currently difficult since the supernova
explosion mechanism is not yet well understood
~\cite[]{Janka.Langanke.ea:2007, Nordhaus.Burrows.etal:2010} and it is
computationally expensive to perform long-time, multidimensional,
systematic studies for different progenitor stars, as would be
desirable in nucleosynthesis studies. Ways to overcome these problems
include using parametric steady-state wind models
~\cite[e.g.][]{Thompson.Burrows.Meyer:2001} and forcing an explosion
by artificially changing neutrinos properties
~\cite[]{Messer.Liebendoerfer.ea:2003, Froehlich.Hauser.ea:2006,
  Froehlich.Martinez-Pinedo.ea:2006, arcones.janka.scheck:2007,
  Fischer.etal:2010}. The evolution of the outflow is rather
independent of the details of the explosion mechanism, but depends
more on the evolution of the neutron star and on the neutrino
emission. Therefore, such approximations are a good basis for
nucleosynthesis studies. Although steady-state wind models cannot
consistently describe hydrodynamical effects (like the reverse shock
and multidimensional instabilities), both approaches agree in the wind
phase~\cite[]{arcones.janka.scheck:2007}.

For the nucleosynthesis studies in this paper, trajectories,
i.e. density and temperature evolutions, from
~\cite{arcones.janka.scheck:2007} were used. These simulations are
based on Newtonian hydrodynamics
~\cite[]{Scheck.Kifonidis.Janka.Mueller:2006, Kifonidis.Plewa.ea:2006}
with general relativistic corrections for the gravitational potential
~\cite[]{Marek.Dimmelmeier.ea:2006} combined with a simplified
neutrino transport treatment
~\cite[]{Scheck.Kifonidis.Janka.Mueller:2006}, which is
computationally very efficient and reproduces the results of Boltzmann
transport simulations qualitatively. The neutrino spectra are assumed
to follow Fermi-Dirac distributions with spectral temperatures
different from the local matter temperature in general.  The central
part ($\rho \gtrsim10^{13} \, \mathrm{g/cm}^3$) of the proto-neutron
star is replaced by a Lagrangian inner boundary placed below the
neutrinosphere. This reduces the computational time and is justified
in part due to the uncertainties in the high-density EoS.  The
evolution of the inner boundary is parametrized by its radius and
neutrino luminosity. The latter is chosen such that an explosion
energy around $10^{51}\mathrm{erg}$ is obtained. The neutrino
luminosity is constant during the first second after bounce and
follows a power law decrease afterwards. We use following models from
~\cite{arcones.janka.scheck:2007}: M10-l1-r1 for a 10$M_{\odot}$
progenitor star (hereafter referred as 10M model), M15-l1-r1 (15M
model) and M15-l1-r6 (15M(s) model) both for a 15$M_{\odot}$, and
M25-l5-r4 for a 25$M_{\odot}$ (25M model). The proto-neutron star
radius and neutrino luminosity evolution are similar for models M10,
M15, and M25. Model M15(s) has a significant different proto-neutron
star radius contraction but same neutrino luminosity. While the
evolution of the inner boundary in model M15 leads to a more compact
neutron star (approximately following the behaviour of the EoS from
~\cite{Lattimer.Swesty:1991}), model M15(s) reproduces a neutron star
with a larger final radius (similar to the EoS from
~\cite{Shen.Toki.ea:1998a}). The different compactness of the neutron
star implies changes in its radius and mass that directly affect wind
parameters (Eqs.~(\ref{eq:QW_S})--(\ref{eq:QW_tau})).

\begin{figure*}[!htb]
    \includegraphics[width=8cm]{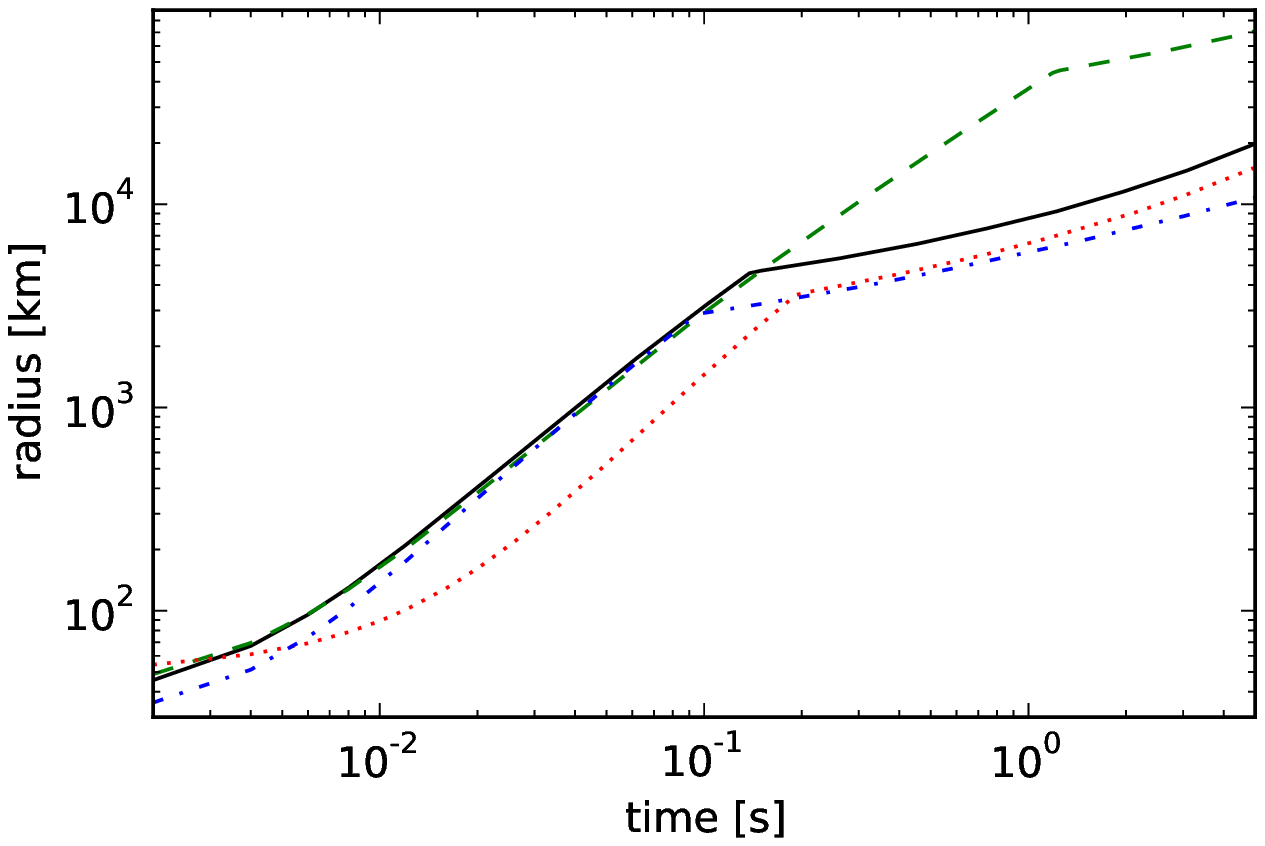}%
    \includegraphics[width=8cm]{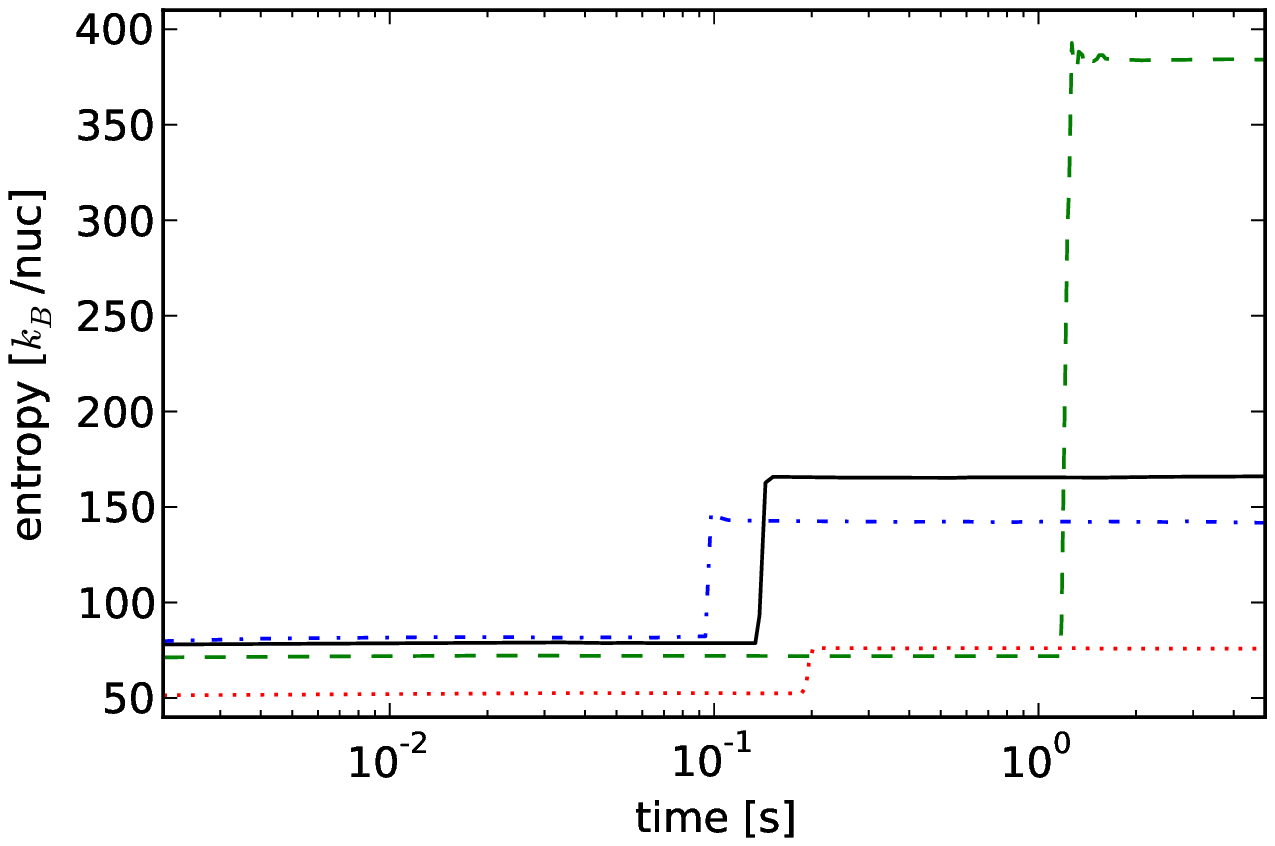}\\
    \includegraphics[width=8cm]{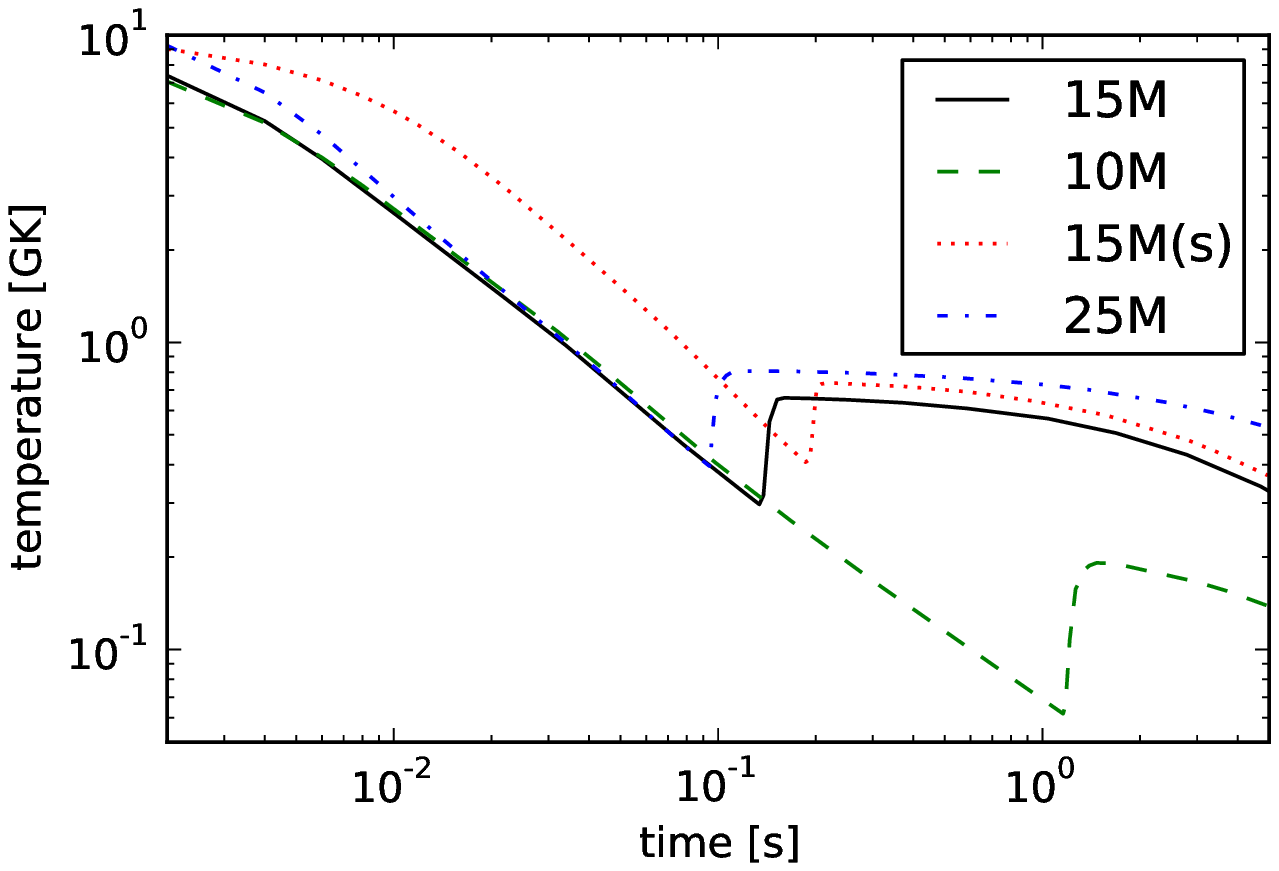}%
    \includegraphics[width=8cm]{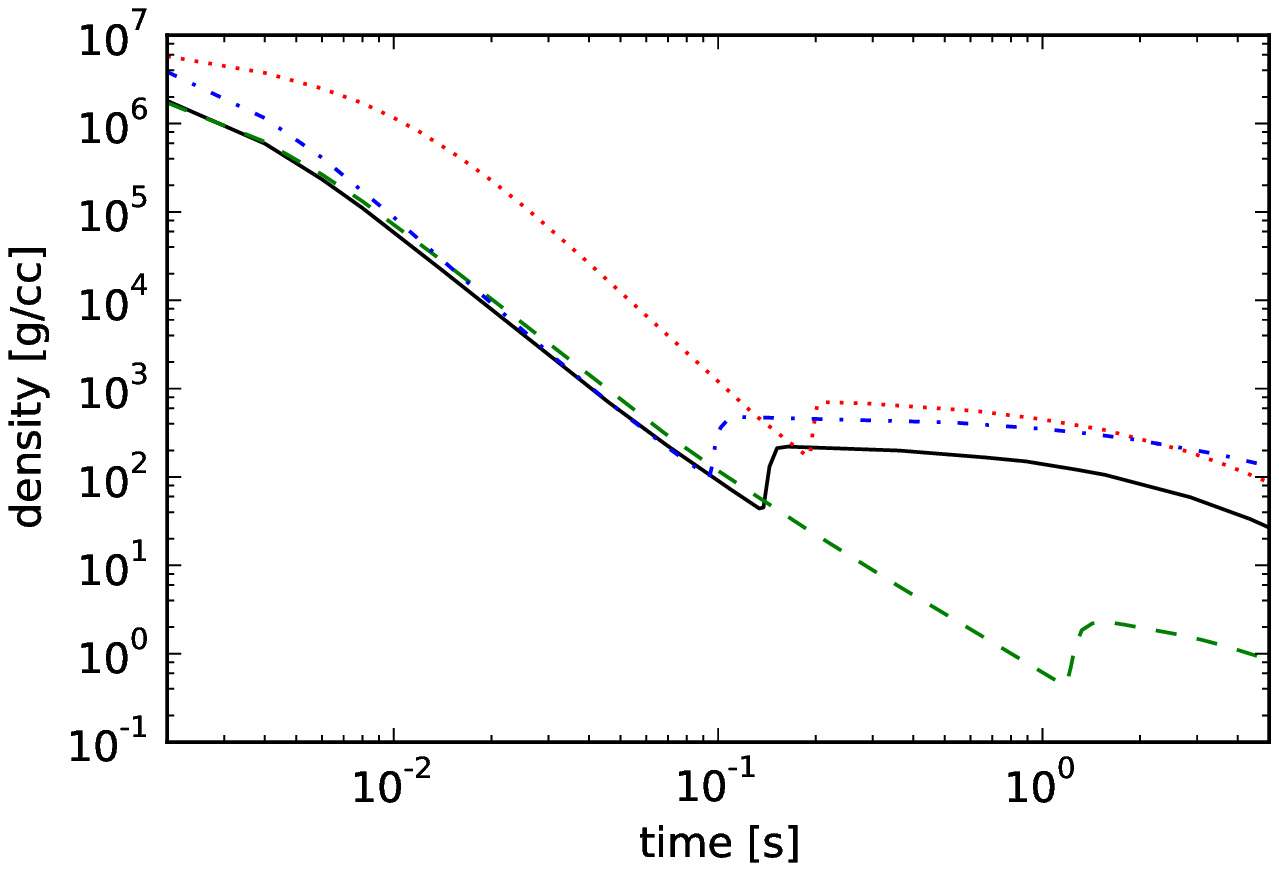}
  \caption{Evolution of radius, entropy, temperature, and density of a
    mass element ejected 5s after bounce for models 10M,
    15M, 15M(s), and 25M.}
  \label{fig:traj}
\end{figure*}

The evolution of radius, entropy, temperature, and density is shown in
Fig.~\ref{fig:traj} for a mass element ejected at 5~s after
bounce. Since the proto-neutron star in models 10M, 15M, and 25M
follows a very similar contraction and neutrino cooling evolution, $S$
and $\tau$ of these models and therefore their density and temperature
evolution are also similar. Their wind profiles have a small
dependence on the mass progenitor. For the most massive progenitor
(25M) the proto-neutron star mass becomes larger (due to the higher
accretion rate) and consequently the wind entropy is also higher than
for the less massive progenitors (see Eq.~(\ref{eq:QW_S})). The impact
of varying the progenitor mass becomes however significant in the
interaction of the wind with the slowly moving, dense supernova
ejecta. This interaction results in a wind termination shock or
reverse shock. The behavior of the reverse shock depends on the wind
properties, but also on the progenitor structure and anisotropies of
the early ejecta (see
~\cite{arcones.janka.scheck:2007,arcones.janka:2010} for more
details). Model 15M(s), with a slow contracting and less compact
proto-neutron star, leads to significantly different wind profiles as
shown Fig.~\ref{fig:traj}. The expansion timescale is longer and the
wind entropy lower than model 15M
(Eqs.~\ref{eq:QW_S}--\ref{eq:QW_tau}).

\subsection{Nucleosynthesis network}
\label{sec:netw}
The evolutions of mass elements ejected between 1~s and 10~s after
bounce are used for our nucleosynthesis calculations. The final
integrated abundances are calculated by adding the abundances from
each mass element weighted by its ejected mass. Each calculation
starts when the temperature decreases below 10~GK. The composition is
calculated initially by assuming nuclear statistical equilibrium (NSE)
for a given initial $Y_e$. The evolution of the composition is
followed using a full reaction network
~\cite[]{Froehlich.Martinez-Pinedo.ea:2006}, which includes 4053
nuclei from H to Hf including both neutron- and proton-rich isotopes.
Reactions with neutral and charged particles are taken from
calculation with the statistical code NON-SMOKER
~\cite[]{Rauscher.Thielemann:2000} and experimental rates are included
~\cite[][NACRE]{Angulo.Arnould.ea:1999} when available. The
theoretical weak interaction rates are the same as in
~\cite{Froehlich.Martinez-Pinedo.ea:2006}. Experimental beta-decay
rates are used when available ~\cite{NuDat2}.

\section{Results}
\label{sec:results}
\begin{figure*}[!htb]
    \includegraphics[width=8cm]{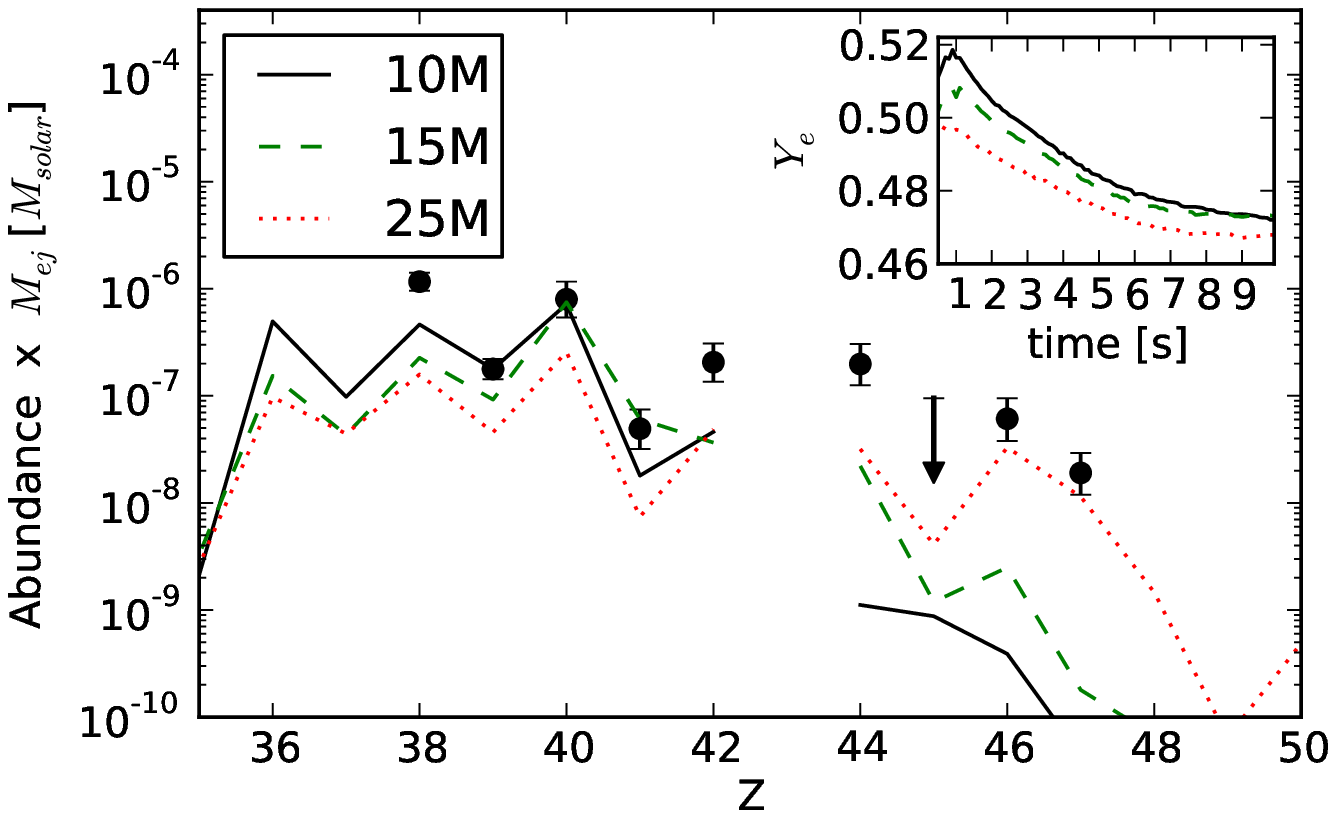}%
    \includegraphics[width=8cm]{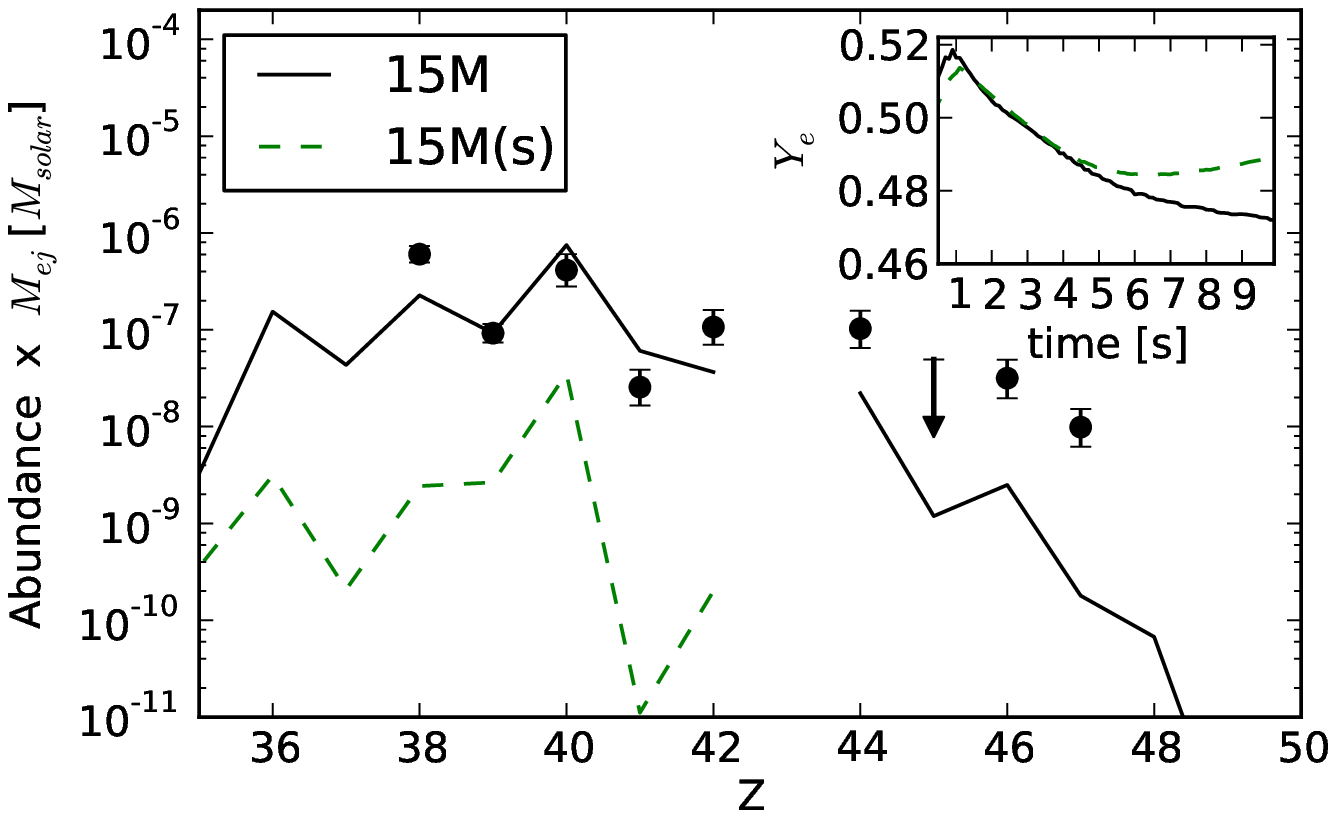}
  \caption{Integrated abundances for the models introduced in
    Sect.~\ref{sec:sn_wind} compared to the LEPP pattern
    ~\cite[]{Qian.Wasserburg:2008} rescaled to $Z=39$. The abundances of
    different progenitors with a similar evolution of the
    proto-neutron star are shown in the left panel, while the right
    panel gives the abundances of the same progenitor with different
    proto-neutron star evolutions.}
  \label{fig:yesim}
\end{figure*}

The integrated abundances for the wind models presented in
Sect.~\ref{sec:sn_wind} are shown in Fig.~\ref{fig:yesim}. The initial
electron fraction and neutrino properties ($L_\nu$ and $\langle
\varepsilon_\nu \rangle$) are taken directly from the wind
simulations.  Based on our simulations, elements are produced up to
$Z=41$ and no heavy r-process elements can be
synthesized. Figure~\ref{fig:yesim} also shows the LEPP elemental
pattern from ~\cite{Qian.Wasserburg:2008} normalized to the calculated
$Z=39$ abundance.  As the proto-neutron star evolution and wind
properties are very similar in models 10M, 15M, and 25M (see
Sect.~\ref{sec:windparameters}), their final abundances are analogous.
The impact of the wind parameters ($S$ and $\tau$) on the abundances
can be seen in the right panel of Fig.~\ref{fig:yesim}. Models 15M and
15M(s) correspond to the same 15~$M_{\odot}$ progenitor, but with
different proto-neutron star evolution (see
Sect.~\ref{sec:sn_wind}). For model 15M the wind expansion is faster
and the entropy becomes higher than in model 15M(s). Model 15M(s) can
produce mainly iron-group nuclei while model 15M leads to heavier
nuclei up to Z=47.

The isotopic production factor indicating the contribution of the
ejected isotope to the solar system abundances is defined as,
\begin{equation}
  P(i)=\sum_j \frac{M_j}{M_{\mathrm{ej}}^{\mathrm{tot}}} \frac{X_j(i)}{X_{\odot}(i)} \, .
  \label{eq:prodfactor}
\end{equation}
The sum goes over every mass element and
$M_{\mathrm{ej}}^{\mathrm{tot}}$ is the total ejected mass obtained
after subtracting the proto-neutron star mass ($M_{\mathrm{ns}}
\approx 1.4 M_\odot$) from the progenitor mass. $X_j(i)$ is the mass
fraction of the isotope $i$ produced in the mass element $j$, and
$X_{\odot,i}$ is the mass fraction of the same isotope $i$ in the
solar system ~\cite[]{Lodders:2003}. Following
~\cite{Qian.Wasserburg:2008}, the solar system was formed from gas
with a contribution of core-collapse supernovae of $M_{\mathrm{gas}}
\approx 3.3 \times 10^{10}M_{\odot}$ ejected during $t\approx
10^{10}$~yr.  The rate of all core-collapse supernovae in the Galaxy
is $R_{\mathrm{SN}} \approx 10^{-2} \mathrm{yr}^{-1}$
~\cite[]{Cappellaro.etal:1999}. This constrains the amount of a given
element $i$ that can be ejected by a single event. If such an event is
representative for all supernovae from different progenitors, this
implies that
\begin{equation}
  \sum_j \frac{M_j}{M_{\mathrm{ej}}^{\mathrm{tot}}} X_j(i) R_{\mathrm{SN}} \,  t \, \langle M_{\mathrm{ej}}^{\mathrm{tot}} 
  \rangle \lesssim M_{\mathrm{gas}}  X_{\odot}(i)\, ,
\label{eq:production}
\end{equation}
has to be satisfied. Here $\langle M_{\mathrm{ej}}^{\mathrm{tot}} \rangle$ is the
average ejected mass, which can be calculated using Salpeter initial
mass function:
\begin{equation}
  \langle M_{\mathrm{prog}} \rangle = 
  \frac{\int_8^{25} m^{-2.5} m \mathrm{d}m}{\int_8^{25} m^{-2.5} \mathrm{d}m} \approx 13 M_\odot \, .
  \label{eq:salpeter}
\end{equation}
Therefore, $\langle M_{\mathrm{ej}}^{\mathrm{tot}} \rangle\ = \langle
M_{\mathrm{prog}} \rangle - M_{\mathrm{ns}} \simeq 11.5 M_{\odot}$,
assuming that all stars in the range $8M_\odot < M< 25M_\odot$
contribute. Because the s-process also contributes to the solar system
inventory, the production factor has to be smaller than the limit
given by Eq.~(\ref{eq:production}):
\begin{equation}
  P(i) \lesssim  \frac{M_{\mathrm{gas}}}
  {R_{\mathrm{SN}} t \langle M_{\mathrm{ej}}^{\mathrm{tot}} \rangle}  \approx 30\,  .
\label{eq:overprod}
\end{equation}
This number is an estimate indicative and comparable with the limit of
~\cite{Mathews.etal:1992} based on the oxygen production in
core-collapse supernovae ($P(i) \lesssim 10$).

\begin{figure*}[!htb]
    \includegraphics[width=8cm]{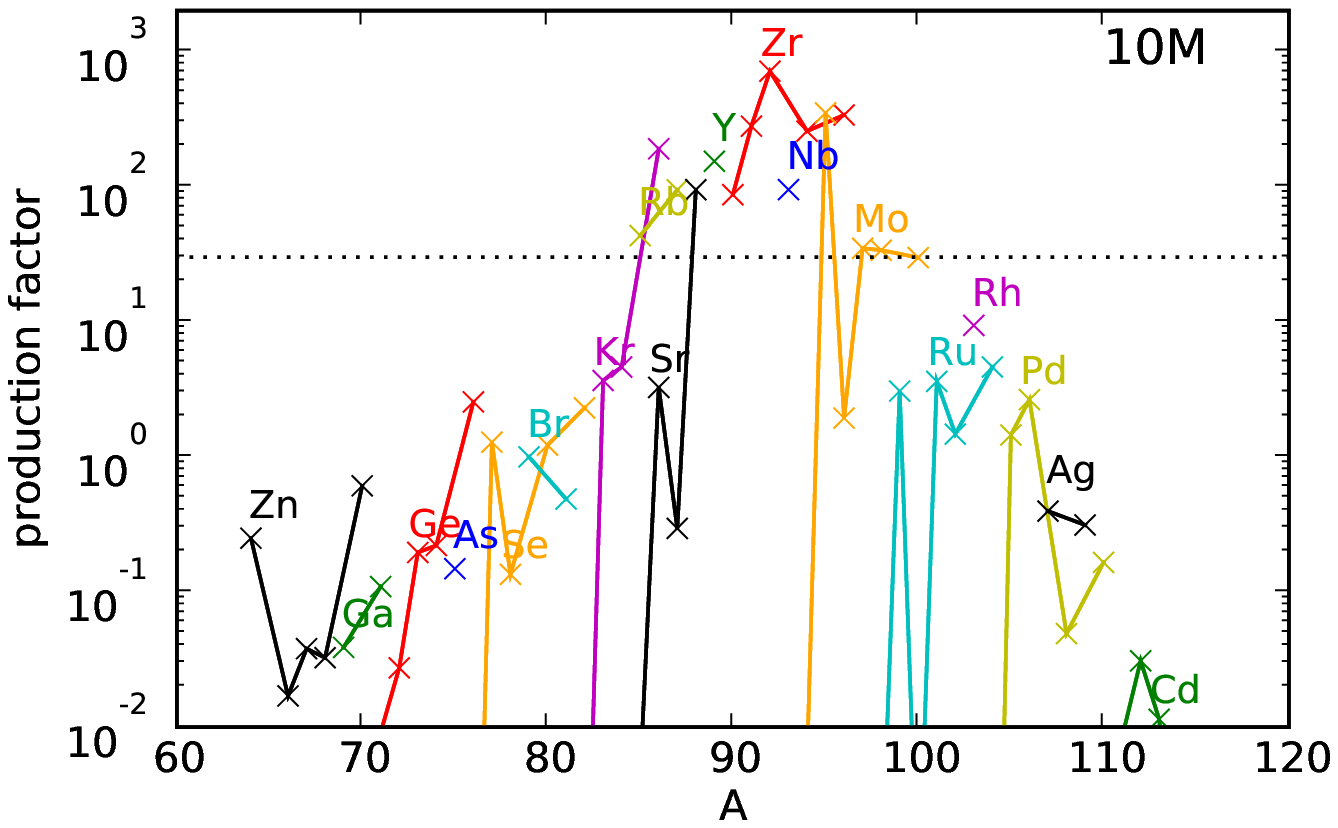}%
    \includegraphics[width=8cm]{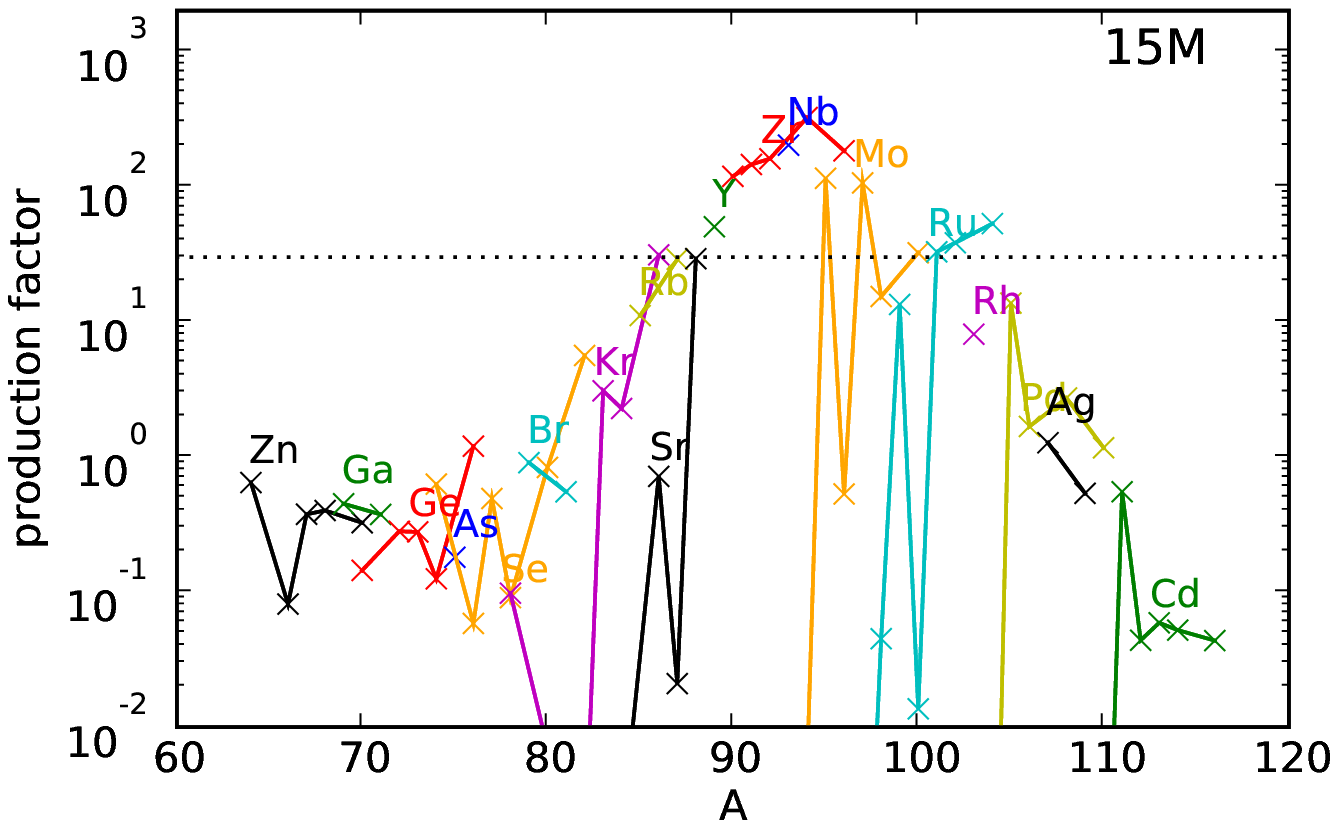}\\
    \includegraphics[width=8cm]{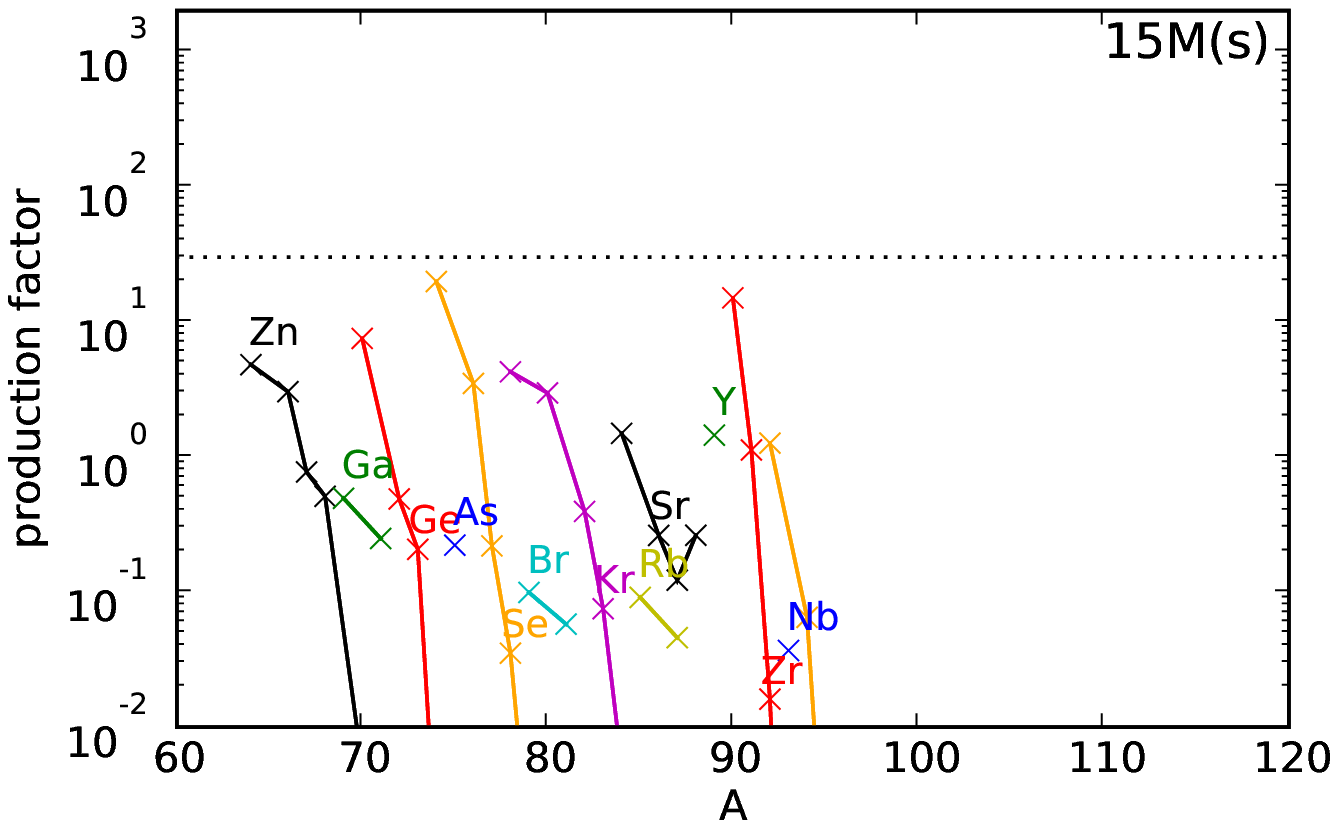}%
    \includegraphics[width=8cm]{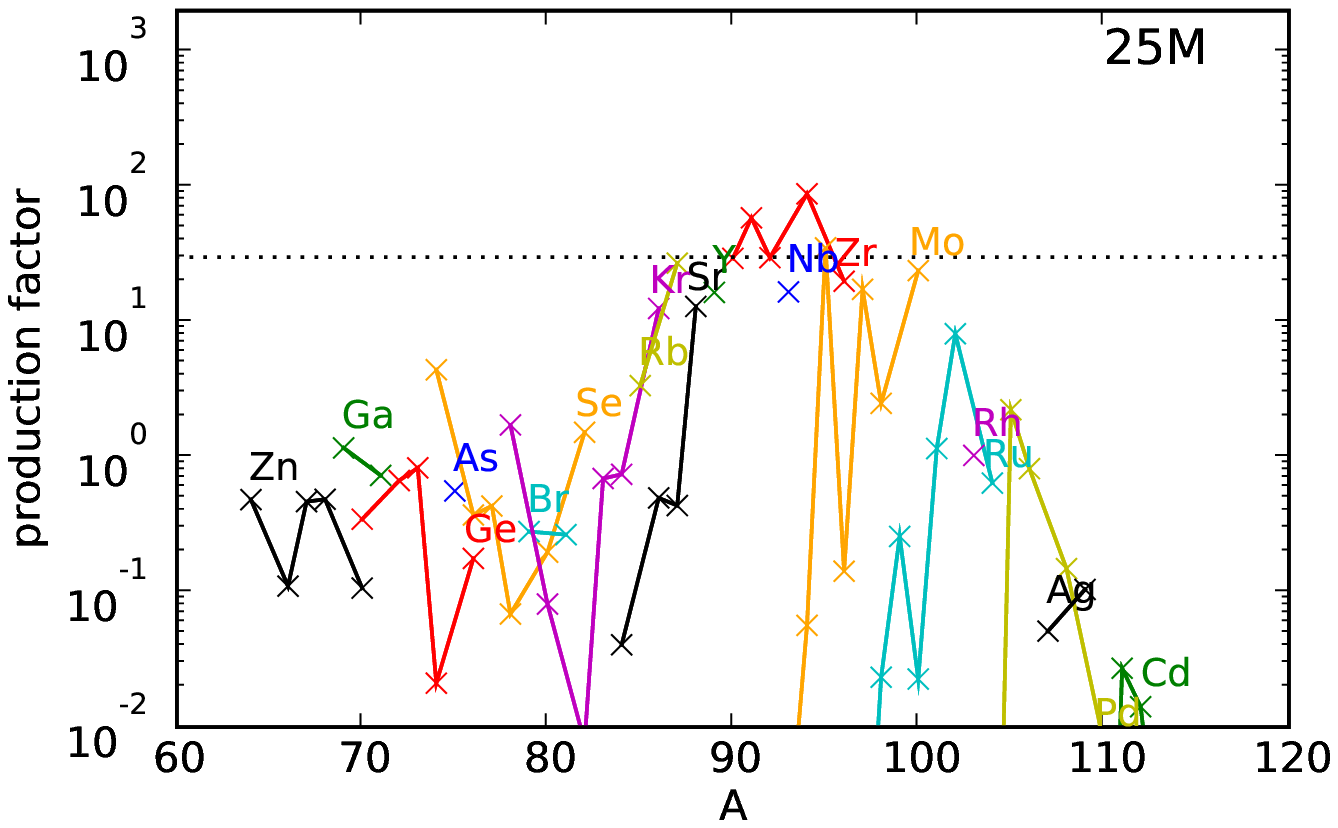}
  \caption{Production factor given by Eq.~(\ref{eq:prodfactor}) for the
    different models shown in Fig.~\ref{fig:yesim}. The horizontal
    dotted line marks the overproduction limit. Isotopes for the same
    element are connected by lines.}
  \label{fig:prodfactor_all}
\end{figure*}

Production factors are shown in Fig.~\ref{fig:prodfactor_all} for all
models. The dotted horizontal line indicates the production limit
given by Eq.~(\ref{eq:overprod}).  Such limit assumes that every
supernova ejects the same amount of matter with the same isotopic
composition. If only a subset of supernova (e.g. those from
progenitors with masses in a determined range, magnetic fields, etc.)
eject those isotopes, the limit will move to larger values.  Note that
only model 15M(s) does not lead to significant amounts of heavy nuclei
because it is producing mainly iron-group nuclei. The overproduction
around $A=90$ is related with the sudden decrease of the alpha
separation energy at magic number $N=50$ ~\cite[see
e.g.,][]{Woosley.Wilson.ea:1994,
  Witti.Janka.Takahashi:1994,Hoffman.Woosley.ea:1996,
  Freiburghaus.Rembges.ea:1999,Hoffman.Mueller.Janka:2008}.  The
production factors reach higher values for low mass progenitors
because the relative contribution of mass ejected is higher.

\subsection{Impact of the electron fraction}
\label{sec:yevar}

The electron fraction is extremely sensitive to details of the
neutrino interactions and transport, and therefore its exact value is
expected to be more uncertain than the value of the other wind
parameters. In this section the impact of $Y_e$ on the production of
LEPP elements is explored. In the following, it is assumed that the
temperature of the electron antineutrinos is $
kT_{\bar{\nu}_e}\approx$~4.5~MeV. For the electron neutrinos,
$kT_{\nu_e} $ is calculated with Eq.~(\ref{eq:ye}) given an initial
$Y_e$. Neutrino luminosities are kept constant as a function of time
for each mass element. For mass element ejected at $t=1$~s after
bounce, $L_{\nu_e}=L_{\bar{\nu}_e}\approx 25 \times 10^{51}\,
\mathrm{erg \,s}^{-1}$. Subsequent mass elements have constant
luminosities $L_{\nu}(t)=L_{\nu}(t=1\mathrm{s})/t^{3/2}$ where $t$ is
the time after core bounce~\cite[]{arcones.janka.scheck:2007}.
Variations of $Y_e$ are linked to neutrino properties which also
affect $S$ and $\tau$. In the extreme cases where the change in $Y_e$
is rather large (i.e., $Y_e=$~0.2~or~0.7), the expansion timescale
changes only by a factor of two (which does not modify its
characteristic timescale of few ms) and the entropy varies by less
than 30\% (Eqs.~(\ref{eq:QW_S})-(\ref{eq:QW_tau})), therefore
justifying our approach of keeping $S$ and $\tau$ as given by the
simulations while varying $Y_e$.

\begin{figure}[!htb]
    \includegraphics[width=8cm]{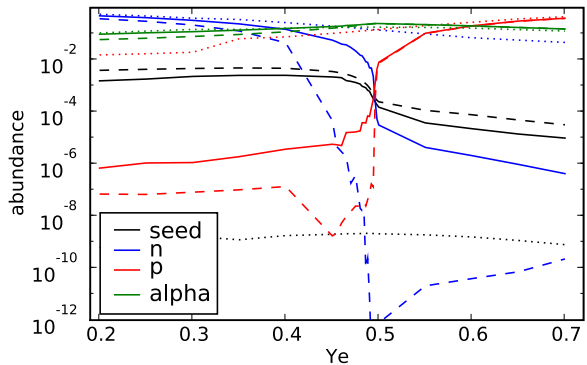}
    \caption{Seed nuclei (i.e., nuclei heavier than $^4$He), neutron,
      proton, and alpha abundances as a function of $Y_e$ based on the
      mass element ejected at 5s after bounce in model 15M. Dotted,
      solid, and dashed lines correspond to temperatures of 8, 5 and 2
      GK, respectively.}
  \label{fig:abund_seed_ye}
\end{figure}

\begin{figure}[!htb]
    \includegraphics[width=8cm]{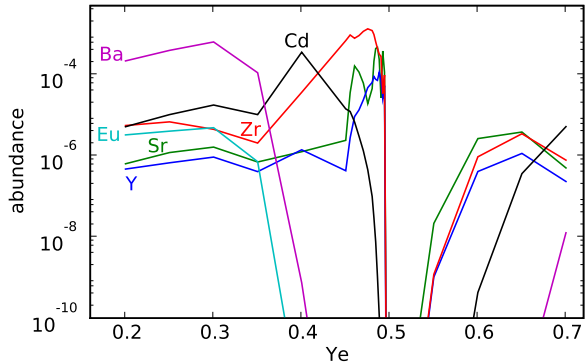}
    \caption{Dependence of the abundances of representative elements
      (Sr, Y, Zr, Cd, Ba and Eu) on the electron fraction. These
      abundances result from a mass element ejected at 5s after
      bounce in model 15M.}
  \label{fig:abund_ye}
\end{figure}

The abundances of neutrons, protons, $\alpha-$particles, and seed
nuclei ($A>4$) as a function of $Y_e$ are shown in
Fig.~\ref{fig:abund_seed_ye} for an expanding mass element ejected 5s
after bounce of model 15M. The abundance at $T=8$~GK consists mainly
of nucleons and $\alpha-$particles. At $T=5$~GK, the abundance of seed
nuclei has significantly increased but $\alpha-$particles still
dominate. Captures of $\alpha-$particles drive the creation of heavier
nuclei until approximately $T=2-3$~GK. The composition at this point
is very important because the subsequent formation of heavier nuclei
depends on the free nucleon-to-seed ratio.  Three different $Y_e$
ranges can be distinguished in
Figs.~\ref{fig:abund_seed_ye}~and~\ref{fig:abund_ye}: proton-rich
($Y_e>0.5 $), rather neutron-rich ($Y_e\lesssim0.4$), and slightly
neutron-rich ($0.4\lesssim Y_e\le0.5 $) conditions.

For proton-rich conditions, heavier nuclei are formed by proton and
$\alpha-$capture reactions for $T>3-5$~GK. If $Y_e$ is large, heavier
elements can be synthesized for lower temperatures due to the increase
of proton and neutron abundances. The amount of neutrons increases
with $Y_e$, i.e. with the number of free protons (see blue dashed line
in Fig.~\ref{fig:abund_seed_ye}). When $Y_e > 0.5$, there are always
more protons available than what can be captured, and only very few
neutrons produced by antineutrino absorption on protons. These
neutrons are immediately captured allowing more matter to bypass
long-lived isotopes (bottlenecks) such as $^{64}$Ge by $(n,p)$
reactions in what is known as the
$\nu$p-process~\cite[]{Froehlich.Hauser.ea:2006,
  Froehlich.Martinez-Pinedo.ea:2006,Pruet.Hoffman.ea:2006,
  Wanajo:2006}.  As the temperature decreases below $\approx 3$~GK a
combination of $(n,p)$, $(n,\gamma)$, and $(p,\gamma)$ reactions carry
the flow to heavier nuclei.  After charged-particle reactions end, the
produced matter decays back to stability, mainly to neutron-deficient
isotopes. Only for early mass elements, the reverse schock becomes
important since its temperature is still in a relevant range ($T
\approx 2$GK) for the $\nu$p-process to occur. The temperature jump at
the reverse shock favors thus the production of heavier nuclei. The
impact of the reverse shock on the $\nu$p-process will be addressed in
future work ~\cite[]{Wanajo.etal:2010, Roberts.etal:2010}.  The
heaviest nucleus that can be reached, depends mainly on the specific
neutrino luminosities and energies, and in less scale, on entropy and
expansion timescale~\cite[see][]{Pruet.Hoffman.ea:2006,
  Wanajo:2006,Martinez-Pinedo.etal:2006}, but as shown in
Fig.~\ref{fig:abund_ye}, LEPP elements can be produced under
proton-rich conditions. As we discussed later, abundances in
proton-rich winds are dominated by p-nuclei.

Extreme neutron-rich conditions are naturally more favorable to reach
heavy r-nuclei. Although these conditions have not been reproduced in
standard supernova ejecta, it may still be possible to obtain them in
scenarios such as explosions with high rotation and magnetic fields
where a jet forms ~\cite[]{Cameron:2001, cameron:2003}, or in quark
nova ~\cite[]{ Ouyed.ea:2002, Jaikumar.ea:2007} where a phase
transition in the neutron star leads to a direct ejection of very
neutron-rich matter. In neutron-rich conditions, the reaction sequence
$^{4}$He($\alpha$n,$\gamma$)$^{9}$Be($\alpha$,n)$^{12}$C
~\cite[]{Woosley.Hoffman:1992, Freiburghaus.Rembges.ea:1999} followed
by $\alpha-$captures quickly increases the amount of seed nuclei to
amounts larger than in proton-rich conditions as shown in
Fig.~\ref{fig:abund_seed_ye}.  Once charged-particle freeze-out
occurs, nuclei are still driven by neutron-captures and $\beta$-decays
before the density gets too low and the created matter $\beta$-decays
towards stability.

When the ejecta are slightly neutron-rich, the
${}^4He$($\alpha$n,$\gamma$) sequence starts to compete with the
triple alpha-reaction and it dominates for $Ye<0.45$. Subsequent
$\alpha$-captures reactions drive the flow following a path close to
or at the valley of beta-stability ~\cite[]{Woosley.Hoffman:1992,
  Witti.Janka.Takahashi:1994, Freiburghaus.Rembges.ea:1999}.  At
closed $N=50$ neutron shell, the alpha separation energy
decreases~\cite[]{Moeller.Nix.Kratz:1997} preventing $\alpha$-captures
to continue with the formation of heavier nuclei. This leads to an
accumulation of matter around $^{88}$Sr, $^{89}$Y,
$^{90}$Zr~\cite[]{Woosley.Hoffman:1992, Witti.Janka.Takahashi:1994,
  Freiburghaus.Rembges.ea:1999}.  When the temperature drops below
$\approx 3$~GK, the Coulomb barrier produces the end of
$\alpha$-reactions and the material captures neutrons if the
neutron-to-seed ratio is still high enough.  When the matter is
symmetric ($Y_e=0.5$) only elements up to the iron-group are formed.

\subsection{LEPP  and $Y_e$ evolution}
\label{sec:lepp}
For the evolution of the initial electron fraction, we use a
parametric approach \cite[similar to][]{Wanajo:2007} with the aim of
fitting the abundances observed in LEPP-enriched UMP stars. In the
following, model 15M is used unless otherwise specified.

\subsubsection{Proton-rich winds}
\label{sec:prich}
\begin{figure}[!htb]
  \includegraphics[width=8cm]{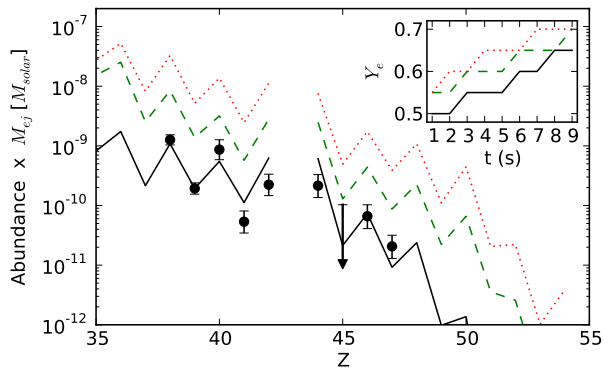}
  \caption{Integrated abundances in model 15M using the three
    different $Y_e$ evolutions shown in the inset. Same colors and
    line styles are used for the abundances and $Y_e$ evolutions. The
    LEPP pattern has been rescaled to fit the solid line abundances.}
  \label{fig:prich_n}
\end{figure}

\begin{figure}[!htb]
  \includegraphics[width=8cm]{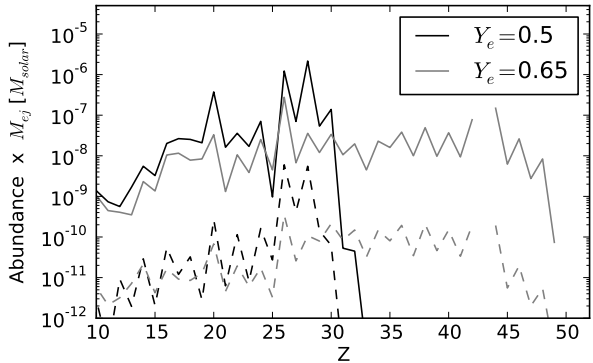}
  \caption{Abundances for the 15M trajectories ejected at 1~s (solid
    lines) and 9~s (dashed lines) after bounce assuming that the
    initial electron fraction is $Y_e=0.5$ (black lines) and
    $Y_e=0.65$ (grey lines).}
  \label{fig:prich1s}
\end{figure}

The integrated abundances corresponding to three different $Y_e$
evolutions are shown in Fig.~\ref{fig:prich_n}. These evolutions are
motivated by the recent results of ~\cite{Huedepohl.ea:2010} for the
explosion and wind evolution of a 8.8~$M_\odot$ stellar progenitor
that shows an increasing $Y_e$ as a function of time.  The abundance
pattern in the LEPP region $38 < Z < 47$ changes only slightly for the
different cases and fits observations within uncertainties. The
robustness of the abundance pattern can be attributed to the
consistency in the $(p,\gamma)$ and $(n,p)$ reaction path for the mass
elements responsable for the abundances in the LEPP region. Most of
those elements are produced when $Y_e>0.6$. Such robustness provides a
natural explanation for the consistency of the suggested UMP star
abundance LEPP pattern mentioned in
~\cite{Montes.etal:2007}. Figure~\ref{fig:prich1s} shows the
similarity of the abundance pattern for $Y_e=0.65$ for two different
mass elements ejected at early and late times. The only significant
difference is the amount of mass ejected in each case.

The main contribution to the integrated abundances come from mass
elements ejected during the first seconds after bounce due to their
large mass outflow. Late mass elements eject less mass but produce
heavier elements because of the higher $Y_e$. These trends are shown
in Fig.~\ref{fig:prich1s} by the mass-weighted abundances at 1~s and
9~s after bounce for two different initial $Y_e$. For $Y_e=0.5$ mainly
iron-group nuclei are produced. LEPP elements can be synthesized for
$Y_e=0.65$ as antineutrino absorption on free protons make possible
$(n,p)$ reactions. As high $Y_e$ values are reached only by mass
elements emitted at late times (small inset in
Fig.~\ref{fig:prich_n}), LEPP abundances are weighted by a lower mass
ejection of the shell. Depending on the $Y_e$ of the early mass
elements, the amount of iron-group elements can change by orders of
magnitude leading to large variations of the ratio between iron-group
and LEPP elements.

While the LEPP elements are most probably produced during the
neutrino-driven wind, the synthesis of iron-group nuclei occurs mainly
in explosive nucleosynthesis when the supernova shock disrupts the
stellar envelope ~\cite[]{Woosley.Heger:2007}. Therefore, the
production of the iron-group elements in the neutrino-driven wind
should be negligible. Although in our simulations iron-group elements
are produced mostly when $Y_e\approx 0.5$, a higher $Y_e$ combined
with a lower neutrino luminosity may also produce them. In our model
15M, the abundances of iron-group elements (in particular, Z=29)
compared to observed abundances of HD~122563~\cite[]{Westin.etal:2000}
can constrain the $Y_e$ of the early trajectories. We find that the
$Y_e$ evolution has to follow either a rapid increase so the amount of
iron-group elements ejected stays low, or it has to start already at
$Y_e \gtrsim 0.55$ so there is only a minor production of these
elements. The evolution represented by the solid line in
Fig.~\ref{fig:prich_n} produces too much iron-group nuclei relative to
LEPP elements as the $Y_e$ stays close to 0.5 for several seconds.
The evolutions represented by dotted and dashed lines are consistent
with observations as the abundances of the iron-group elements are
less than 10\% of the observed abundances when normalizing to Z=39.

\begin{figure*}[!htb]
    \includegraphics[width=8cm]{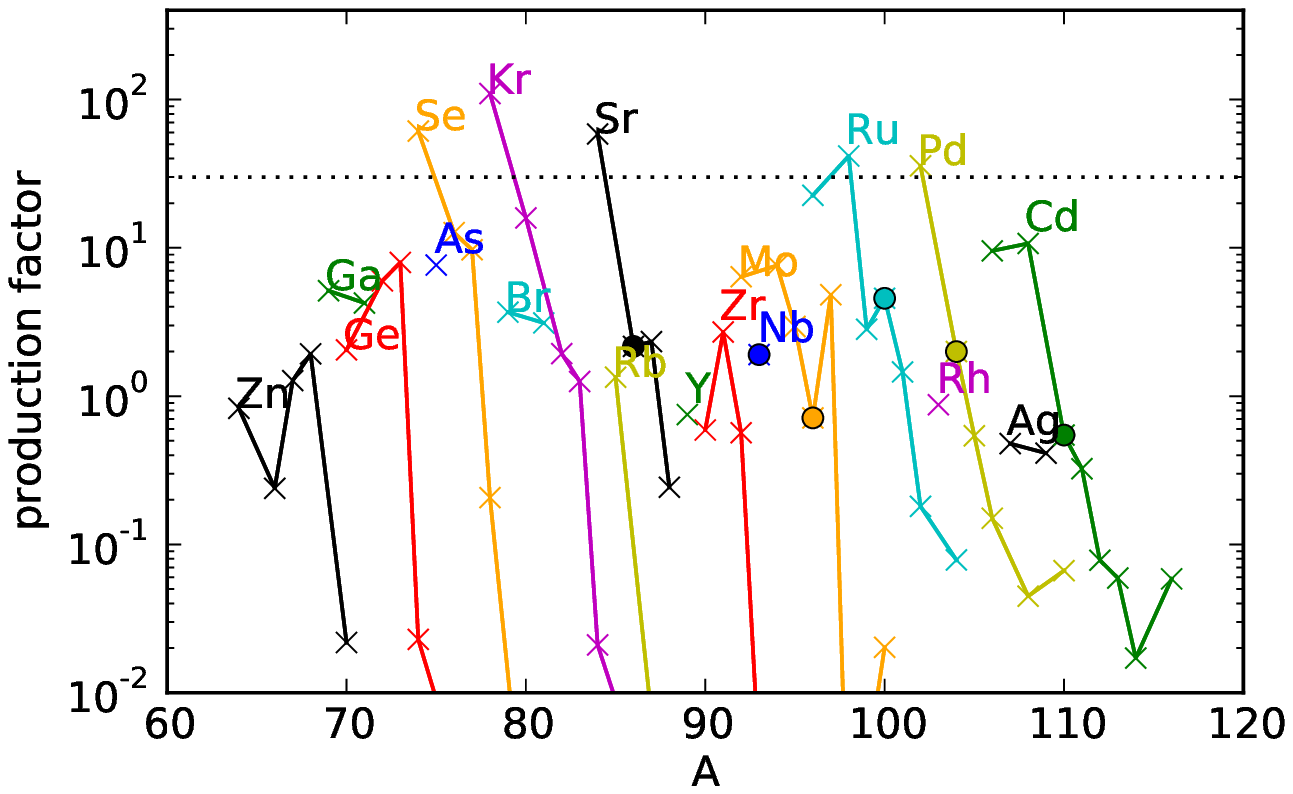}%
    \includegraphics[width=8cm]{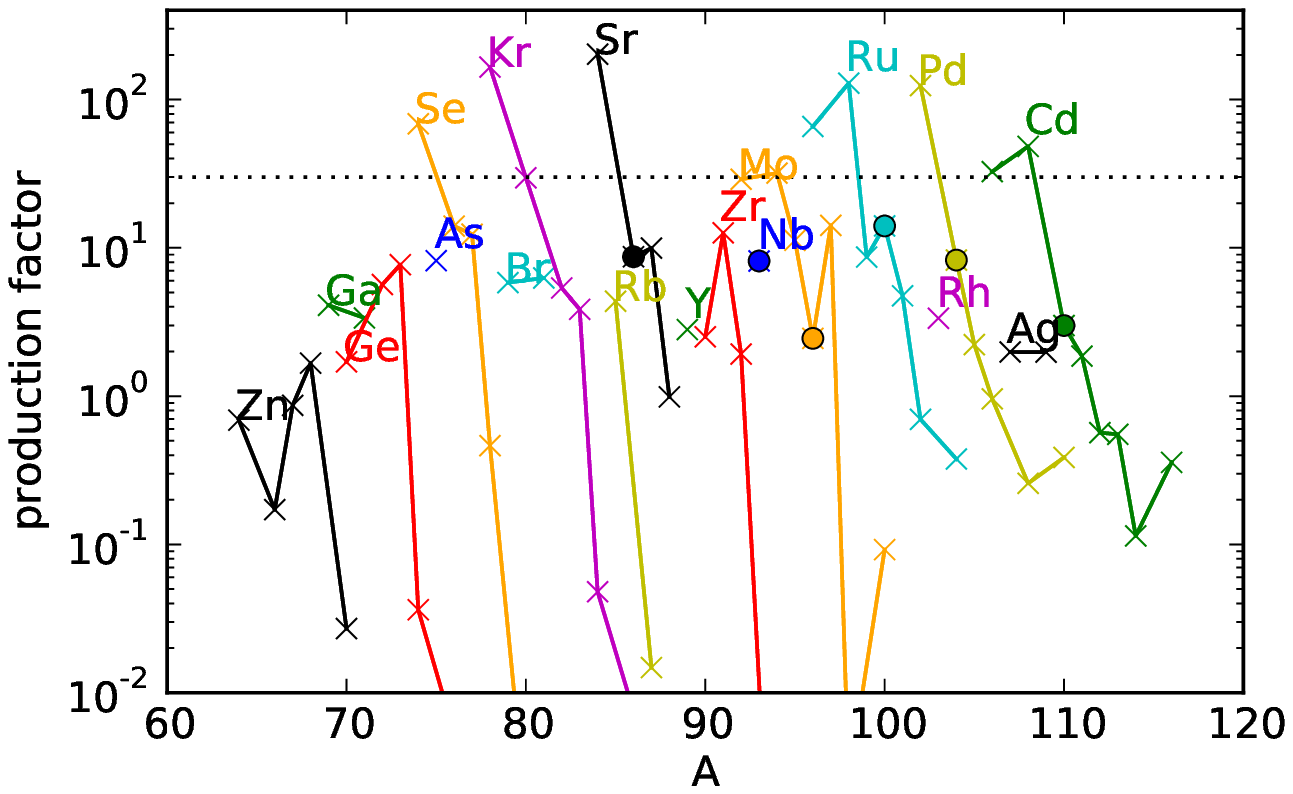}
    \caption{Production factors corresponding to the elemental
      abundances presented in Fig.~\ref{fig:prich_n} by the dashed and
      dotted lines are shown in the left and right panels,
      respectively.  Almost only neutron-deficient isotopes (p-nuclei)
      are synthesized. The isotopes marked with circles are not
      produced in enough quantities in
      ~\cite{Travaglio.Gallino.ea:2004}.}
  \label{fig:prich_overprod}
\end{figure*}

Production factors corresponding to different proton-rich evolutions
are shown in Fig.~\ref{fig:prich_overprod}. To gain some insight into
the absolute abundances of LEPP elements, one may compare their
abundances with the measured values of $\alpha$-elements that are the
dominant products of core-collapse supernovae. For instance, Mg should
take the value given by Eq.~(\ref{eq:overprod}) since it is a typical
supernova product. Assuming HD~122563 is representative for all
supernovae and [Mg/Y]=0.9~\cite[]{Honda.etal:2004,Honda.etal:2006},
one would expect the production factor of Y (single isotope) to be
around~3. Figure~\ref{fig:prich_overprod} shows that $P($Y$)$ has
roughly the right value. Nevertheless, that some isotopes are over the
production limit given by Eq.~(\ref{eq:overprod}) means that either
not all supernovae in the range $8 M_\odot< M < 25 M_\odot$ reach the
same proton-rich conditions or that the ejecta from our spherically
symmetric, parametric model cannot be extrapolated to every supernova
ejecta. If the abundances observed in HD~122563 are not representative
of a typical supernovae ejecta then Y (or other LEPP elements) could
be produced in different amounts.  We find that abundances from models
10M and 25M behave very similar to 15M. For model 15M(s) is not
possible to produce LEPP elements even when $Y_e \sim 0.65$ due to the
low entropy ~\cite[]{Pruet.Hoffman.ea:2006}.  In addition, most of the
abundances are on the proton-rich side of the valley of beta-stability
(neutron-deficient isotopes). If the LEPP were responsible for the
isotopic underabundances reported by
~\cite{Travaglio.Gallino.ea:2004}, proton-rich winds would not be
enough to account for this contribution.

\begin{figure}[!htb]
  \includegraphics[width=8cm]{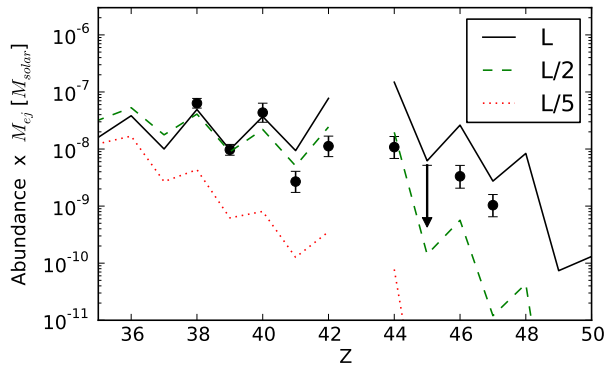}
  \caption{Abundance patterns obtained for the mass element ejected
    1~s after bounce with $Y_e =0.65$ and different neutrino
    luminosity. Solid line shows the abundances using the luminosity
    from \cite{arcones.janka.scheck:2007}. For the other cases the
    luminosity has been reduced by different factors as indicated in
    the label.}
  \label{fig:prichL}
\end{figure}

The nucleosynthesis dependance on neutrino luminosity is shown in
Fig.~\ref{fig:prichL} for a trajectory ejected 1~s after bounce with
$Y_e=0.65$. The neutrino luminosity obtained in
\cite{arcones.janka.scheck:2007} has been reduced by different
factors. When the neutrino luminosity decreases the wind entropy has
to increase (Eq.~\ref{eq:QW_S}). For our test, we do not consider this
dependence and keep the entropy (and other wind parameters)
unchanged. Although, this is not fully consistent, it allows us to
investigate the influence of the neutrino luminosity on the abundances
separately from the effect of other wind parameters.  We find that a
reduction of the luminosity by a factor 2 (dashed line, ``L/2'') leads
to the production of only $Z < 44$ elements.  A factor of 5 reduction
is enough to inhibit the production of any LEPP elements, as the flow
can barely pass through the bottleneck at $^{64}$Ge. We note that
previous nucleosynthesis studies based on the explosion model of
~\cite{Kitaura06} do not produce LEPP elements
~\cite[]{Hoffman.Mueller.Janka:2008, Wanajo.Nomoto.ea:2009} because
the (anti)neutrino luminosity of this low mass progenitor
($M=8.8M_\odot$) is rather small and the expansion is very
fast. Therefore, there are not enough neutrinos during sufficient time
to get a successful $\nu$p-process.  The neutrino luminosities
obtained for heavier mass progenitors, as in our case and in
~\cite{Pruet.Hoffman.ea:2006} for a 15~$M_\odot$ progenitor, show that
with higher luminosities the $\nu$p-process can occur in the supernova
outflows. Although we have shown that with the calculated
(anti)neutrino luminosities it is possible to obtain the LEPP pattern,
the neutrino luminosities from hydrodynamical simulations have large
uncertainty and prevents us to draw definite conclusions.

\subsubsection{Neutron-rich winds}
\label{sec:nrich}
\begin{figure}[!htb]
    \includegraphics[width=8cm]{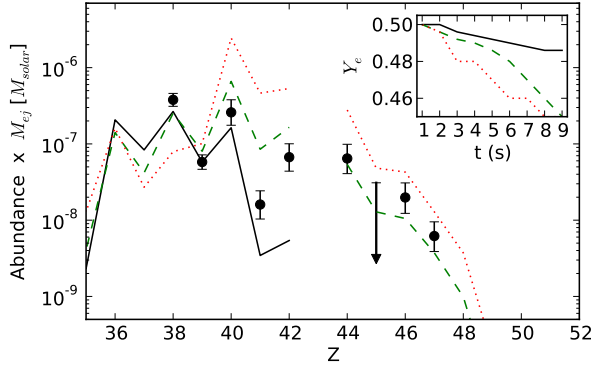}
    \caption{Same as Fig.~\ref{fig:prich_n} but for neutron-rich $Y_e$
      evolutions.}
  \label{fig:nrich_evol}
\end{figure}

Figure~\ref{fig:nrich_evol} shows integrated abundances for three
possible neutron-rich $Y_e$ evolutions (inset of the same
Figure). When the $Y_e$ decreases from $0.5$ to $0.486$, only $Z\le40$
are produced (similar to Fig.~\ref{fig:yesim}).  If $Y_e$ decreases to
$0.45$, elements up to $Z=47$ can be produced. Contrary to proton-rich
conditions, different $Y_e$ evolutions lead to significant changes in
the pattern. This sensitivity in the final abundances is due to the
abrupt drop of the neutron abundance near $Y_e=0.5$ (shown in
Fig.~\ref{fig:abund_seed_ye}). This rapid change in neutron abundance
is accompanied by a non-linear change in LEPP abundances when
$0.4<Y_e<0.5$ (shown in Fig.~\ref{fig:abund_ye}).  Moreover, mass
elements ejected at slightly different times (and therefore with small
variations of the entropy and the expansion timescales) show a
substantial difference on the neutron-to-seed ratio and subsequently
on the abundances, making again difficult to get a robust
pattern. Such strong dependence of the final abundances on the wind
parameters is also found for the other models introduced in
Sect.~\ref{sec:results}. In model 15M(s) the electron fraction has to
be reduced down to $Y_e \approx 0.4$ to produce elements up to $Z=47$.

\begin{figure*}[!htb]
    \includegraphics[width=8cm]{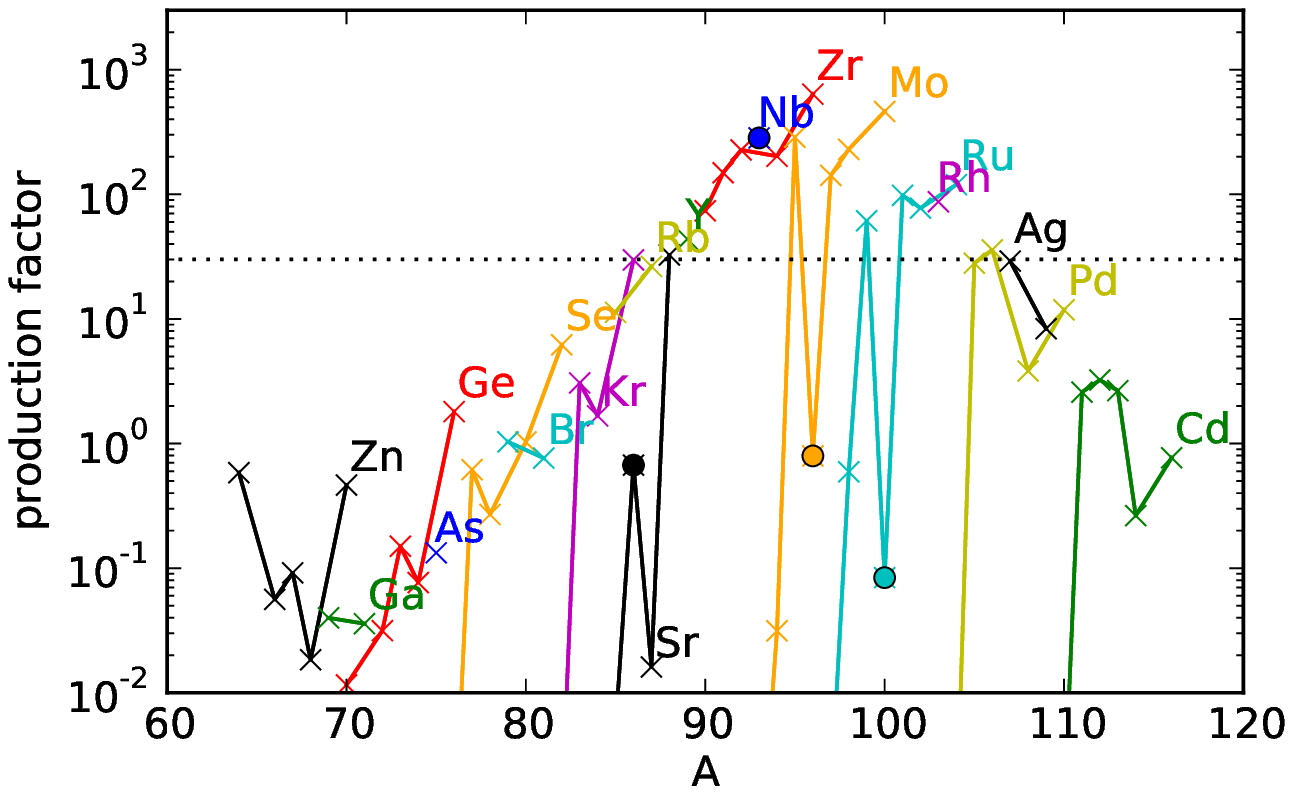}%
    \includegraphics[width=8cm]{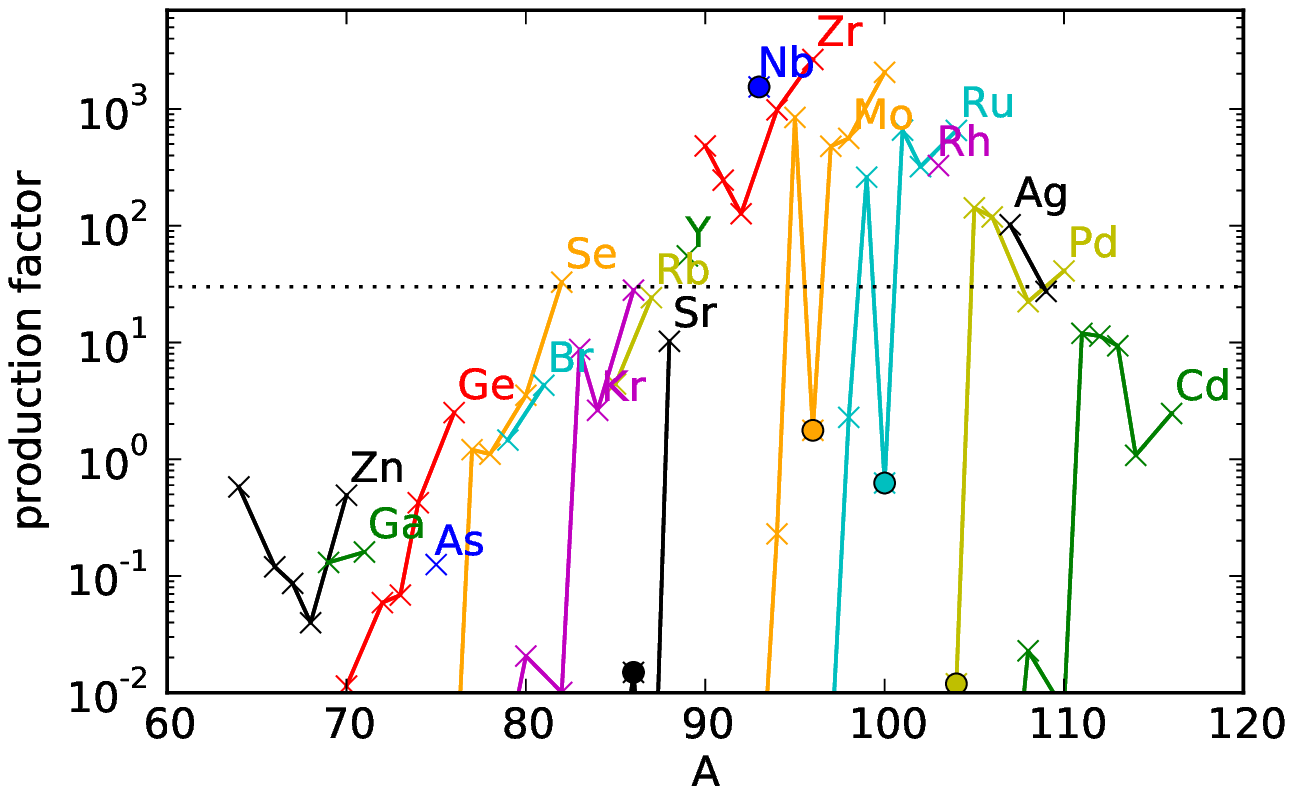}
    \caption{Production factors for two evolutions of the electron
      fraction: left panel correspond to green dashed line and right
      panel to the red dotted line in Fig.~\ref{fig:nrich_evol}. The
      dotted horizontal line represents the upper limit above which
      isotopes are overproduced (see Sect.~\ref{sec:yevar}).}
  \label{fig:nrich}
\end{figure*}

Production factors of the abundances shown by dashed and dotted lines
in Fig.~\ref{fig:nrich_evol} are presented in the left and right
panels of Fig.~\ref{fig:nrich}, respectively. Since in neutron-rich
winds ($0.45 \lesssim Y_e < 0.5$) the amount of heavy nuclei produced
is higher than in the proton-rich winds (Fig.~\ref{fig:abund_ye}), the
overproduction around A~=~90 (related to the neutron magic number
N~=~50, ~\cite{Hoffman.Woosley.ea:1996}) becomes a constraint. Similar
to the conclusion reached for the proton-rich conditions, this
suggests that not all mass ejected in neutrino-driven winds in
core-collapse supernovae experience neutron-rich conditions. Previous
results ~\cite[see e.g.,][]{Woosley.Hoffman:1992,%
  Woosley.Wilson.ea:1994,Witti.Janka.Takahashi:1994,%
  Hoffman.Woosley.ea:1996,Freiburghaus.Rembges.ea:1999,%
  Hoffman.Mueller.Janka:2008} suggest that only a small amount of the
mass ejected by core-collapse supernovae can be neutron-rich.  It is
possible that the neutron richness of the ejecta depends on progenitor
mass because more massive stars lead to more compact neutron stars and
thus higher neutrino energies, which may favour neutron-rich
conditions.  A different alternative to avoid the overproduction is to
have anisotropic explosions with an electron fraction distribution in
the ejecta that achieves neutron-rich conditions only in some
regions. Recent two-dimensional supernova simulations present
proton-rich ejecta with neutron-rich blobs or pockets that contain
only a small mass ~\cite[]{Wanajo.etal:2010b}.

Further investigations are necessary to conclude which of the
possibilities addressed here can explain the observations of LEPP
elements in UMP stars. Based on spherical symmetric simulations we
have showed that it is possible to reproduce the LEPP elemental
pattern under different wind conditions. Multidimensional simulations
will determine the effect of anisotropic $Y_e$ distribution on the
wind nucleosynthesis.  Such multidimensional simulations combined with
exploratory studies like ours will provide constraints for the
specific conditions (e.g., $Y_e$ evolution) and for the overall
contribution of neutron- and proton-rich winds to the solar system
inventory. Using our results, it is hard to obtain the solar LEPP
pattern ~\cite[]{Travaglio.Gallino.ea:2004,Montes.etal:2007} by
combining neutron- and proton-rich conditions. Moreover, Galactic
chemical evolution models are necessary before a definite conclusion
can be drawn to the exact contribution of core-collapse supernovae to
the solar systems abundances.

\section{Conclusions and summary}
\label{sec:conclusions}
We have performed nucleosynthesis calculations based on spherically
symmetric supernova simulations ~\cite[]{arcones.janka.scheck:2007}
which include efficient neutrino transport and follow the evolution of
the ejecta from a few milliseconds to several seconds after bounce
during the neutrino-driven wind phase. The integrated nucleosynthesis
was analyzed for different stellar progenitors with masses of
10,~15,~and~25~$M_\odot$.  We have found that light element primary
process abundances can be produced under realistic conditions in the
neutrino-wind phase.  No heavy r-process elements can be synthesized
under those conditions~\cite[see
also][]{Roberts.etal:2010,Wanajo.etal:2010b}.

Our results indicate that the nucleosynthesis occurring in
neutrino-driven winds does not depend significantly on the progenitor,
but rather on the proto-neutron star evolution which in turn directly
affects nucleosynthesis-relevant wind parameters
~\cite[]{Qian.Woosley:1996} such as entropy, expansion timescale, and
electron fraction. The electron fraction depends on neutrino
properties which are determined by the still uncertain neutrino
interactions and transport. Since the calculation of the electron
fraction remains challenging ~\cite[]{Huedepohl.ea:2010} the impact of
the electron fraction on the production of LEPP elements was
studied. The LEPP pattern obtained by ~\cite{Qian.Wasserburg:2007,
  Montes.etal:2007} was reproduced for different evolutions of the
electron fraction towards proton- and neutron-rich conditions.

The LEPP pattern reproduced in proton-rich winds was found to be
robust under small variations in the evolution wind
parameters. Nevertheless, elements heavier than iron-group nuclei can
only be produced when the neutrino fluxes are high enough to allow for
a successful $\nu$p-process. The amount of heavy elements ejected is
rather low since most of the outflow matter consists of protons and
$\alpha$-particles. Although the elemental abundances nicely reproduce
the observed LEPP pattern, mainly neutron-deficient isotopes are
produced \cite[see also][]{Froehlich.Hauser.ea:2006,
  Froehlich.Martinez-Pinedo.ea:2006, Wanajo.etal:2010}. Therefore,
proton-rich conditions can explain the LEPP elements observed in UMP
stars but not in the missing isotopic underabundances in the solar
system~\cite[]{Travaglio.Gallino.ea:2004}.

When the electron fraction is assumed to evolve towards neutron-rich
conditions, the LEPP pattern can be also reproduced but it is not
robust under small variations of the wind parameters.  We find an
overproduction around $A \sim 90$ that was already pointed out in
previous nucleosynthesis studies ~\cite[]{Woosley.Hoffman:1992,
  Woosley.Wilson.ea:1994, Witti.Janka.Takahashi:1994,
  Freiburghaus.Rembges.ea:1999, Hoffman.Mueller.Janka:2008}. This
overproduction and the fact that most recent supernova simulations
~\cite[]{Fischer.etal:2010, Huedepohl.ea:2010} favor proton-rich winds
could suggest that neutron-rich winds are rare events. Scenarios that
could explain the ejection of only a small amount of neutron-rich
material include 1) very small pockets with neutron-rich material that
may appear in multidimensional
simulations~\cite[]{Wanajo.etal:2010b}; 2) neutron-rich
conditions that can only be obtained at very late
times when the amount of ejected mass is very small; 3) there is only
a small subset of all supernovae that can develop neutron-rich winds.

The isotopic underabundances reported by
\cite[]{Travaglio.Gallino.ea:2004} cannot be reproduced based in our
models without overproducing other isotopes. Evolving from proton to
neutron conditions in the wind does not resolve the overproduction,
which could be a hint of the limitation of our models. However, this
can also suggest deficiencies in the s-process model used by
\cite[]{Travaglio.Gallino.ea:2004}. We cannot conclude whether the
solar and the UMP stars LEPP abundances are produced by the same light
element primary process.

Observation of isotopic abundances in UMP stars would constrain the
evolution of the electron fraction and thus of the neutrino properties
in supernovae. In addition more work is necessary in improving
long-time supernova models (e.g., multidimensional simulations, more
accurate treatment of neutrino reactions and transport) and also the
experimental and theoretical nuclear reactions relevant in the wind
nucleosynthesis before definite conclusions can be reached.

\begin{acknowledgements}
  We thank H.~Th.~Janka, K.~Langanke, G.~Mart\'inez-Pinedo, H.~Schatz,
  and F.~K.~Thielemann for stimulating discussions.  We are grateful to
  G.~Mart\'inez-Pinedo for providing us the nucleosynthesis
  network. A.~Arcones acknowledges support of the Deutsche
  Forschungsgemeinschaft through contract SFB 634, ExtreMe Matter
  Institute EMMI, and Swiss National Science Foundation. F.~Montes is
  supported by NSF grants PHY 08-22648 (Joint Institute for Nuclear
  Astrophysics) and PHY 01-10253.
\end{acknowledgements}


\end{document}